\documentclass[prd,twocolumn,superscriptaddress,floatfix,nofootinbib]{revtex4}
\usepackage{color}

\usepackage{bm}
\usepackage{graphicx}
\usepackage{mathrsfs}
\newcommand{\clf}{\mathcal{F}}

\def\reff@jnl#1{{\rm#1\ }}

\def\aj{\reff@jnl{AJ}}                  
\def\araa{\reff@jnl{ARA\&A}}            
\def\apjl{\reff@jnl{Astrophys.\ J.\ Lett.}}  
\def\apjs{\reff@jnl{Astrophys.\ J.\ Suppl.\ Ser.}}              
\def\ao{\reff@jnl{Appl.Optics}}         
\def\apss{\reff@jnl{Ap\&SS}}            
\def\aap{\reff@jnl{A\&A}}               
\def\aapr{\reff@jnl{A\&A~Rev.\ }}         
\def\aaps{\reff@jnl{A\&AS}}             
\def\azh{\reff@jnl{AZh}}                
\def\baas{\reff@jnl{BAAS}}              
\def\jcap{\reff@jnl{J.\ Cosmol.\ Astropart.\ Phys.\ }}	
\def\jhep{\reff@jnl{JHEP}}		
\def\jrasc{\reff@jnl{JRASC}}            
\def\memras{\reff@jnl{MmRAS}}           
\def\mnras{\reff@jnl{Mon.\ Not.\ R.\ Astron.\ Soc.}}            
\def\npb{\reff@jnl{Nulc.\ Phys.\ B}}        
\def\plb{\reff@jnl{Phys.\ Lett.\ B}} 	
\def\pr{\reff@jnl{Phys.\ Rev.\ }}        	
\def\pra{\reff@jnl{Phys.\ Rev.\ A}}         
\def\prb{\reff@jnl{Phys.\ Rev.\ B}}         
\def\prc{\reff@jnl{Phys.\ Rev.\ C}}         
\def\prl{\reff@jnl{Phys.\ Rev.\ Lett.}}      
\def\pasp{\reff@jnl{PASP}}              
\def\pasj{\reff@jnl{PASJ}}              
\def\qjras{\reff@jnl{QJRAS}}            
\def\skytel{\reff@jnl{S\&T}}            
\def\solphys{\reff@jnl{Solar~Phys.}}    
\def\sovast{\reff@jnl{Soviet~Ast.}}     
\def\ssr{\reff@jnl{Space~Sci.\ Rev.}}     
\def\zap{\reff@jnl{ZAp}}                
\def\nat{\reff@jnl{Nature}}             

\begin{document}

\title{What can be learned from the lensed cosmic microwave background
$B$-mode polarization power spectrum?}

\author{Sarah Smith}
\author{Anthony Challinor}
\affiliation{Astrophysics Group, Cavendish Laboratory,
J J Thomson Avenue, Cambridge CB3 0HE, United Kingdom}
\author{Gra\c{c}a Rocha}
\affiliation{Astrophysics Group, Cavendish Laboratory, J J Thomson Avenue, Cambridge CB3 0HE, United Kingdom}
\affiliation{Observational Cosmology, California Institute
of Technology, Pasadena CA 91125, USA}
\affiliation{Jet Propulsion Laboratory, Pasadena CA 91109, USA}

\begin{abstract}
The effect of weak gravitational lensing on the cosmic microwave
background (CMB) temperature anisotropies and polarization will provide
access to cosmological information that cannot be obtained from the
primary anisotropies alone. We compare the information content
of the lensed $B$-mode polarization power spectrum, properly accounting
for the non-Gaussian correlations between the power on different scales,
with that of the unlensed CMB fields and the lensing potential. The latter
represent the products of an (idealised) optimal analysis that exploits the
lens-induced non-Gaussianity to reconstruct the fields. Compressing the
non-Gaussian lensed CMB into power spectra is wasteful and
leaves a tight degeneracy between the equation of state of dark energy and
neutrino mass that is much stronger than in the more optimal analysis.
Despite this, a power spectrum analysis will be a useful first step in analysing
future $B$-mode polarization data. For this reason, we also consider how to
extract accurate parameter constraints from the lensed $B$-mode power spectrum.
We show with simulations that for cosmic-variance-limited
measurements of the lensed $B$-mode power, including the non-Gaussian
correlations in existing likelihood approximations gives biased parameter
results. We develop a more refined likelihood approximation that performs
significantly better. This new approximation should also be of more general
interest in the wider context of parameter estimation from Gaussian CMB data.
\end{abstract}


\maketitle

%
%

\section{Introduction}
\label{sec:intro}

The goal of many current and forthcoming cosmic microwave background (CMB)
experiments is to measure the polarization of the CMB with increasing
accuracy (e.g.\ QUaD~\cite{POL/Chu++03}, BICEP~\cite{POL/Kea++03},
EBEX~\cite{POL/Oxl++04},
QUIET\footnote{\protect \url{http://quiet.uchicago.edu/}} and
Clover~\cite{POL/Tay++??}).
There are several motivations for this: polarization
measurements are an important test of the consistency of the cosmological
model, as well containing new information, and can break parameter 
degeneracies which are present if only the CMB temperature power spectrum is
analysed~\cite{DA/ZSS97}. A key motivation is the fact that the
curl-like $B$-mode polarization is not generated directly from scalar
perturbations at last scattering, and thus can potentially reveal the
presence of a primordial gravitational wave background, if it exists at a
high enough level~\cite{POL/KKS97a,POL/S+Z97}.
A detection of primordial large-angle $B$-mode polarization would thus
provide a measure of the expansion rate and hence energy scale of
inflation. Apart from the reionization bump around $\ell \simeq 5$,
which depends on the optical depth to last scattering, 
the gravitational wave signal peaks at multipoles $\ell \sim 100$,
corresponding to the angle subtended by the horizon size at last scattering,
and so large area surveys (a few hundred deg$^2$) will be required to detect
it. On smaller scales, the dominant contribution to $B$-mode polarization is
expected to be the weak lensing of $E$-modes by large scale structure
along the line of sight~\cite{LEN/Z+S98}.

The lensing signal is sensitive to the gravitational potential, and
hence the clustering of matter, mostly at redshifts $\alt 10$~\cite{LEN/Sel96}.
(At the peak of the lensing deflection power spectrum at multipoles
$\ell \sim 40$, more than 95\% of the power arises from $z < 10$.)
Lensing is therefore sensitive to parameters that have little direct effect
on the pattern of primordial fluctuations, but
affect the late-time growth of large-scale structure.
The lensing signal can thus provide us with information
which would otherwise be absent from the CMB~\cite{LEN/M+S97, LEN/S+E99},
such as the properties of dark energy and sub-eV neutrino
masses~\cite{LEN/Hu02}.
However, the lensed $B$-modes are
significantly non-Gaussian, and this has been shown to degrade
markedly the amount of information present in the $B$-mode power
spectrum~\cite{LEN/SHK04}.
The non-Gaussianity introduced by weak lensing is also present to a much lesser
extent in the CMB temperature anisotropies and $E$-mode polarization signal. 
It has recently been shown that neglecting this 
non-Gaussianity does not significantly bias parameter constraints from
Planck-quality data, although the effect of weak lensing on the
power spectra cannot be neglected~\cite{LEN/Lew05}.

In the present paper we consider more fully the question of what parameter
information is contained in the lensed $B$-mode power spectrum
and how to analyse future spectral data reliably.
We concentrate on the $B$-mode spectrum here for the following reason.
The expectation is that
for future experiments with sufficient signal-to-noise to image the
lens-induced $B$-modes, most of the additional
parameter information from the lensing effect on the CMB power spectra will
come from the $B$-modes. Although
the information content of the $B$-mode power spectrum is relatively
more affected by the non-Gaussianity than the temperature ($T$) and
$E$-mode polarization, that latter must contend with the cosmic
variance of the dominant primary (unlensed) contributions.
It is well known that for Gaussian fields the observed power spectrum is a
sufficient statistic for parameter estimation under ideal survey conditions,
but that this is not the case for non-Gaussian fields such as the lensed
CMB. Compressing the lensed CMB fields to their power spectra is wasteful
leading to a loss of cosmological information. A more optimal analysis
would be to use the non-Gaussianity to reconstruct an estimate of the
lensing deflection field and the unlensed CMB
fields~\cite{LEN/Hu01,LEN/H+O02}, or to work directly from the correct
non-Gaussian likelihood function~\cite{LEN/H+S03,LEN/H+S03b}.
While such analyses are a worthy goal to strive for, they will likely be
difficult to implement in practice in the presence of real-world
complications such as inhomogeneous noise, complex survey geometries
and foreground residuals.
A simpler, and probably more robust method to deal with near-future
lensed data is to work directly with the lensed fields in a conventional
power spectrum analysis. 

This paper is organised as follows. In Section~\ref{sec:params} we review how
$B$-mode polarization is generated by the weak lensing of $E$-modes, and
consider the cosmological parameters which can be constrained using this
information, and the degeneracies between them. In Section~\ref{sec:gaussian}
we estimate these parameters from simulated lensed data
under the assumption that the data are Gaussian, in order to show directly that 
this leads to false conclusions.
In order to account for the non-Gaussianity we
need to understand the correlation between the measured power on different
scales, and we calculate this in Section~\ref{sec:cov} following
Ref.~\cite{LEN/SHK04}. In Section \ref{sec:ng-like} 
we incorporate the non-Gaussian covariance in existing likelihoods
used for the analysis of (Gaussian) CMB data but show that this too leads
to biased results in simulations. The reason for this deficiency
appears to arise from inaccuracies in the current likelihood
approximations when applied to data with few degrees of freedom
(which extends to higher multipoles for the lensed $B$-modes than for Gaussian
fields). We introduce a more accurate likelihood function in
Section~\ref{sec:new-like} which we show to perform much better on simulated
lensed data. An appendix provides further details of the derivation
of this new likelihood.

%
%

\section{Cosmological parameters}
\label{sec:params}
\begin{figure}
\includegraphics[width = \columnwidth]{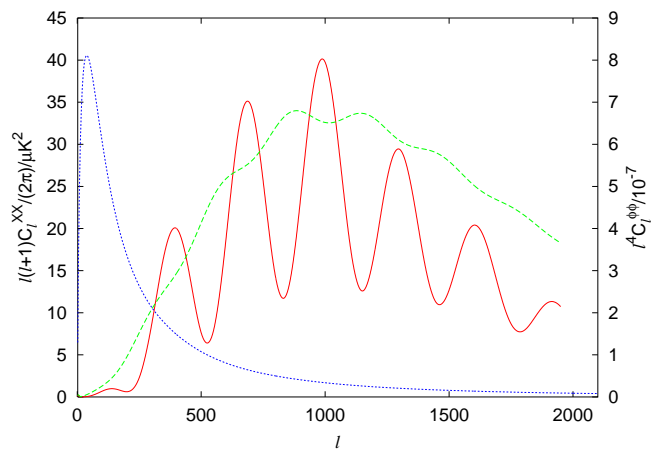}
\caption{(Colour online) Unlensed $\tilde{E}\tilde{E}$ power spectrum (red solid line) and power
spectrum of lensing potential (blue dotted line) which together generate the lensed
$BB$ power spectrum (shown by green dashed line, multiplied by a factor of 400).}
\label{fig:E_B_phi}
\end{figure}

The action of lensing on the part of the temperature anisotropy and
polarization that is generated at recombination is a simple re-mapping
that in the flat-sky approximation can be written as
\begin{equation}
X(\bm{n}) = \tilde{X}(\bm{n}+\nabla \phi).
\label{eq:born}
\end{equation}
Here $\tilde{X}(\bm{n})$ is the unlensed variable ($T$, or the Stokes
parameters $Q$ and $U$),
$X(\bm{n})$ the lensed value and $\phi$ the lensing deflection field.
On the sphere, we need to take care to interpret correctly the meaning
of $\nabla \phi$ \cite{LEN/C+C02} and to take into account the
position-dependence of the co-ordinate basis for polarization.
The root-mean-square lensing deflection angle $\sim 3\, \text{arcmin}$
and so the lensing action on CMB fields well above this scale can be
evaluated with the gradient approximation:
\begin{equation}
X(\bm{n})  \approx \tilde{X}(\bm{n}) + \nabla \tilde{X} \cdot \nabla \phi.
\label{eq:grad-approx}
\end{equation}
Within this approximation, and in the flat-sky limit, we can express
the lensed $B$-mode power
spectrum as a convolution of the unlensed $E$-mode power spectrum with the
power spectrum of the lensing potential $\phi$ \cite{LEN/SHK04}:
\begin{equation}
C_\ell^{BB} =
\int \frac {\mathrm{d^2}\bm{\ell}'}{(2\pi)^2}
C_{\ell'}^{\tilde{E}\tilde{E}} C_{|\bm{\ell} - \bm{\ell}'|}^{\phi\phi}
W^2(\bm{\ell}, \bm{\ell}'),
\label{eq:C^BB}
\end{equation}
where function $W$ is given by
\begin{equation}
W(\bm{\ell}, \bm{\ell}') = \bm{\ell}'\bm{\cdot} (\bm{\ell}-\bm{\ell}')
\sin 2(\phi_{\bm{\ell}}-\phi_{\bm{\ell}'}),
\label{eq:W}
\end{equation}
with
$\bm{\ell} = (\ell \cos \phi_{\bm{\ell}}, \ell \sin \phi_{\bm{\ell}})$.
This is illustrated in Fig.~\ref{fig:E_B_phi}, which shows the unlensed
$E$-mode and lensing potential power spectra, together with the resulting
lensed $B$-mode power spectrum. The power spectra were calculated using
{\sc camb}\footnote{\protect \url{http://camb.info/}}
which includes corrections for curved sky effects and the breakdown of the
gradient approximation~\cite{LEN/C+L05}.

The lensed $B$-mode power spectrum is thus affected by 
cosmological parameters which affect either the primary (unlensed) CMB power
spectrum $C_\ell^{\tilde{E}\tilde{E}}$ or the lensing power
$C_\ell^{\phi\phi}$. Many of these parameters can be well constrained by the
temperature and $E$-mode power spectra, and the lens-induced $B$-modes do not
contribute significantly to improving these constraints. Here we concentrate
on those parameters that are degenerate with respect to the unlensed
CMB spectra. As is well known, the unlensed spectra can provide tight
constraints on the physical densities in baryons and cold dark matter,
the angular diameter distance to last scattering, the primordial power
spectrum, and the optical depth to reionization~\cite{DA/E+B99}.
Neutrinos with masses
below $\sim 0.3\,\text{eV}$, as implied by current analyses of large-scale
structure and the Ly-$\alpha$ forest~\cite{COS/Sel++05}, are relativistic at
recombination and
the only effect of their mass on the unlensed CMB is via the angular diameter
distance,
$d_A$, and a small large-scale contribution to the late integrated
Sachs-Wolfe (ISW) effect~\cite{DA/Eis++99}.
Similarly, dark energy parameters, such
as the current energy density $\Omega_\Lambda h^2$, and equation of state
$w$ are felt only through $d_A$ and the ISW effect. In the inflation-inspired
flat models considered here, the parameters $\Omega_\Lambda h^2$, $w$, and
the current density in massive neutrinos $\Omega_\nu h^2$ (or equivalently
the neutrino masses) are highly degenerate with respect to the unlensed
CMB, even for cosmic-variance-limited observations. They are collectively
constrained only by $d_A$, which is now accurately determined to be
$13.7 \pm 0.5\, \text{Gpc}$ from the CMB alone~\cite{COS/Spe++03}. Adding
curvature or evolution in the dark-energy sector increases the
dimensionality of the degeneracy. 

While this `geometric' degeneracy can be easily broken with the inclusion of
external data~\cite{DA/E+B99}, it is still worthwhile to consider to what
extent the lensing of the CMB helps given its relatively simple and well-understood
physics. An early analysis of the temperature and $E$-mode
polarization power spectra showed that the weak lensing effect on the
power spectra can go some way to breaking the geometric
degeneracy~\cite{LEN/S+E99} since the late-time evolution of the gravitational
potential responds differently to variations in the parameters that are
geometrically degenerate. Since for multipoles $\ell > 100$ 
the $B$-mode power spectrum is expected to be dominated by weak lensing, this
is potentially a much more sensitive probe of the lensing effect
than the temperature or $E$-mode power since there
is no cosmic variance coming from unlensed $B$-mode polarization.
For a given survey, once
the sensitivity reaches the limit for imaging the lens-induced $B$-modes
(better than $5\, \mu\text{K-arcmin}$) we can expect these
to be most constraining. To reach the ultimate cosmic-variance limit
considered later in this paper, we require sensitivities around
$1\, \mu\text{K-arcmin}$ over a large fraction of the sky. This is similar
to the specifications being discussed for a post-Planck orbital experiment
dedicated to CMB
polarization\footnote{\protect\url{%
http://universe.nasa.gov/program/inflation.html}}.

\begin{figure}
\includegraphics[width = \columnwidth]{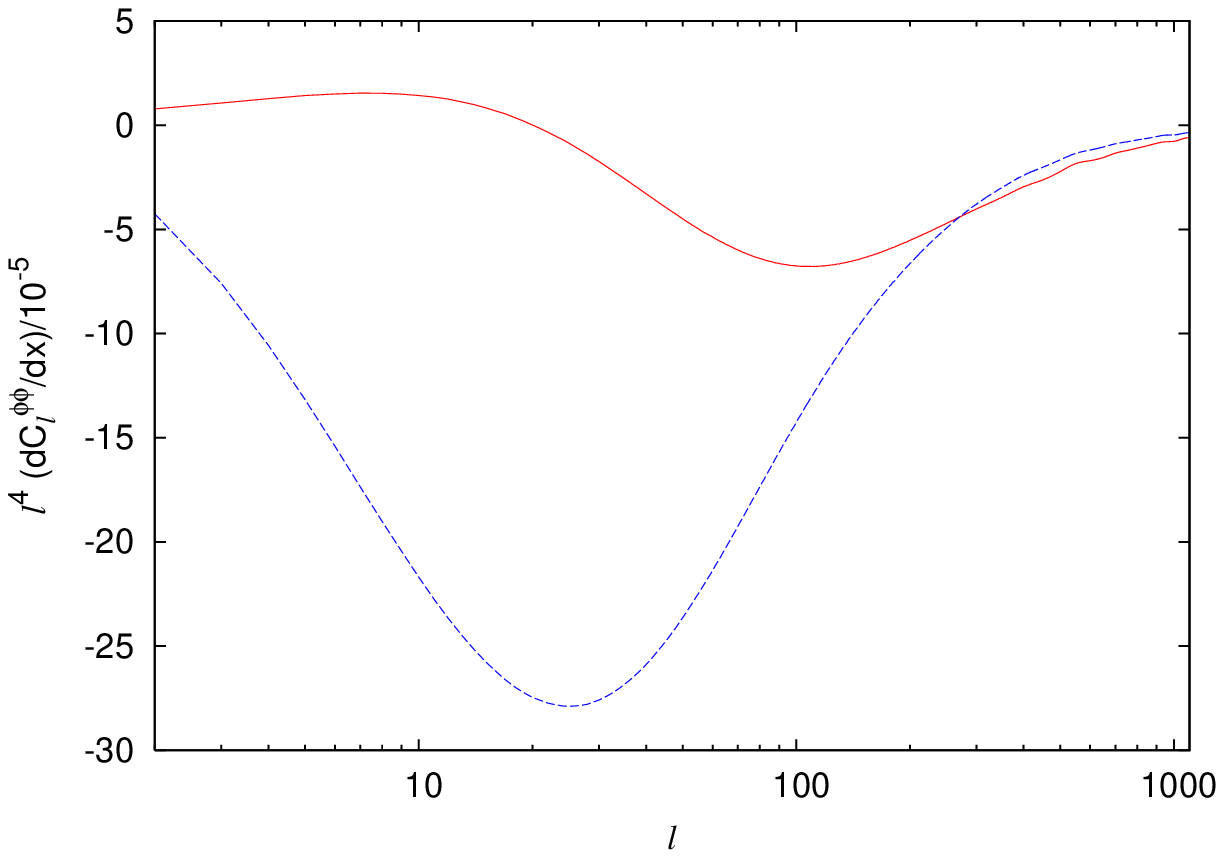}
\includegraphics[width = \columnwidth]{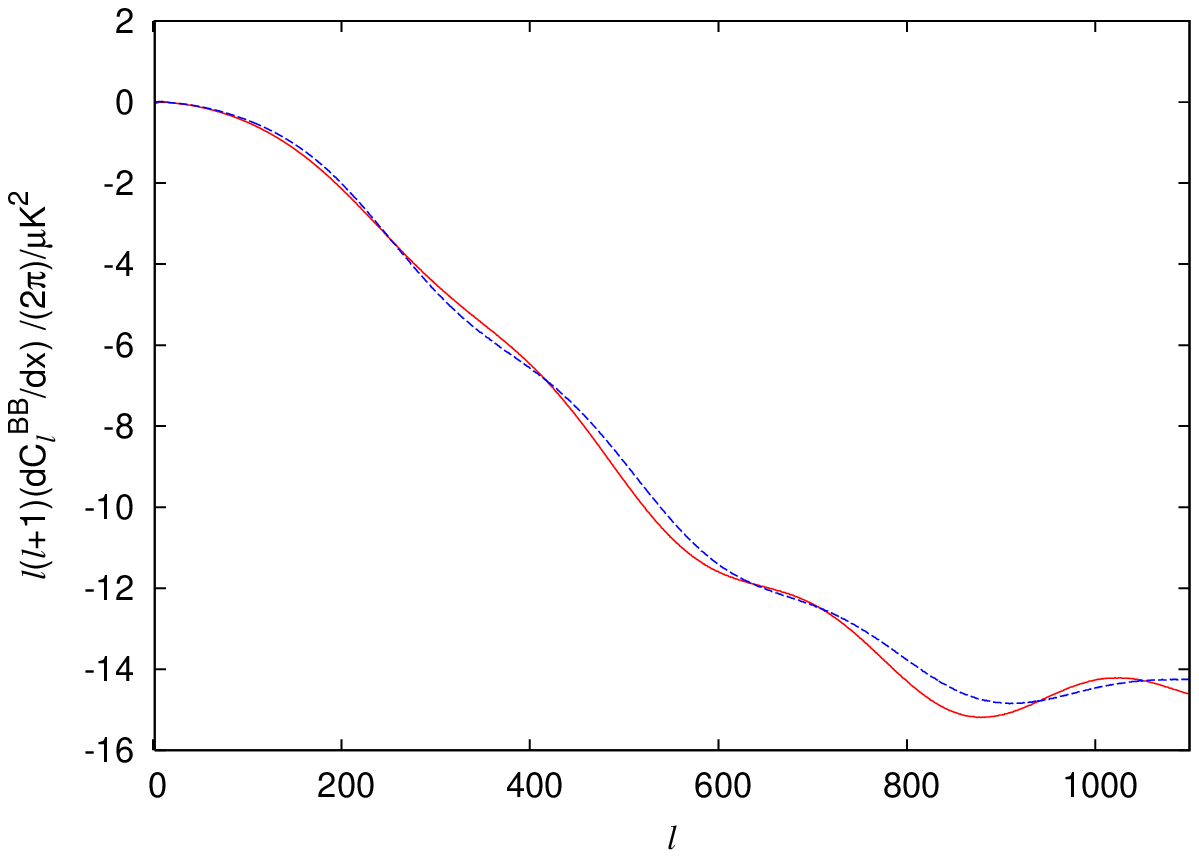}
\caption{(Colour online) Derivatives of the lensing power spectrum
$\ell^4 C_\ell^{\phi\phi}$ (top)
and the lensed $B$-mode power spectrum $\ell(\ell+1) C_\ell^{BB}/(2\pi)$
(bottom),
keeping $\theta_D$ fixed (see text). 
Red solid line: derivatives with respect to $\Omega_\nu h^2$;
blue dashed line: derivatives with respect to $w$ (amplified by a factor of
350 in both plots).}
\label{fig:w_on_deriv}
\end{figure}

In Fig.~\ref{fig:w_on_deriv} we show the derivatives of the lensing
potential and $B$-mode power spectra with respect to $w$ and $\Omega_\nu h^2$
in a flat universe with the physical baryon and cold dark matter
densities held fixed at $\Omega_b h^2=0.024$ and $\Omega_c h^2=0.111$. 
The derivatives are taken around $w = -1$ (a cosmological constant) and
$\Omega_\nu h^2 = 0.004$ with the ratio $\theta_D$ of the sound horizon at
last scattering to the angular diameter distance held fixed. If we
parameterise flat models by $\theta_D$, $w$ and $\Omega_\nu h^2$,
both the Hubble parameter $h$ and $\Omega_\Lambda h^2$ are derived parameters.
Our choice of parameterisation is motivated by the geometric degeneracy:
the unlensed CMB accurately determines all our parameters expect $w$ and
$\Omega_\nu h^2$ which are themselves very poorly constrained in the absence
of external information. The power spectra used to compute these derivatives
were computed with the input unlensed power spectra set to zero above the
maximum multipole value $\ell_{\text{max}} = 2048$ to be consistent with the
analyses of simulations presented in the following sections of this paper.
We have assumed three families of neutrinos with equal masses. Since
$\sum m_\nu = \Omega_\nu h^2 \times 94\, \text{eV}$ when all families
are non-relativistic\footnote{%
We note that $\Omega_\nu h^2$ tends to $1.7 \times 10^{-5}$ in the limit that
the neutrino masses tend to zero (for three families of leptons). It is a good
approximation to take the neutrino mass of a family to be proportional to
its contribution to $\Omega_\nu h^2$ provided that $m_\nu > 0.004\, \text{eV}$.
Since in the November 2004 release of \textsc{camb} that we use 
the (degenerate) masses are calculated as
$\sum m_\nu = \Omega_\nu h^2 \times 94\, \text{eV}$, it is
important to avoid calculating derivatives with respect to $\Omega_\nu h^2$
around $\Omega_\nu h^2 = 0$.
}, the mass in the fiducial model
is $0.13\, \text{eV}$. The total mass is close to the current best
limits from cosmological probes~\cite{COS/Sel++05}. 
Given the measurements of (squared) mass differences
from atmospheric and solar neutrino experiments, our value of
$\Omega_\nu h^2$ is just at the limit where we can assume mass
degeneracy~\cite{COS/E+L05}. We note also that we have not included
any non-linear corrections to the matter power spectrum in computing
the spectra in Fig.~\ref{fig:w_on_deriv} since the
\textsc{halofit}~\cite{COS/Smi++03} fitting employed in \textsc{camb} does not
currently support massive neutrinos or models with $w\neq -1$.
Non-linearities increase the lensing power spectrum beyond
$l \sim 300$, and have an effect $>5\%$ on all scales
for the $B$-mode spectrum, and a much larger effect on small
scales~\cite{LEN/C+L05}. Non-linearities should thus properly be included in any
future data analysis.

The effect of changes in $w$ and $\Omega_\nu h^2$ on the lensing power spectrum
has been discussed in Ref.~\cite{LEN/KKS03}. The main effect of changes
of $w$ is through the change in the expansion rate. An increase in $w$ at
fixed $\theta_D$ (or, equivalently, $d_A$) requires a reduction in
the current dark-energy density and hence $h$. However, increasing
$w$ causes dark energy to dominate earlier and the net effect is an
initial enhancement of the expansion rate over that for $w=-1$; this
causes the gravitational potential to decay earlier and suppresses the
lensing effect. 
Hence the derivative of the lensing potential power spectrum with
respect to $w$ is negative; see Fig.~\ref{fig:w_on_deriv}.
The effect on the potential is almost independent of
scale, but its time dependence causes a larger fractional change in
lensing power on large scales. If we fix the current physical density in
dark energy, the effect of massive neutrinos on the gravitational
potential vanishes on large scales (larger than the Jeans length
of the neutrinos when they first become non-relativistic) since the
increased expansion rate is offset by the neutrino clustering. The
effect on the lensing power spectrum thus also vanishes at large $\ell$.
On scales below the neutrino Jeans length today the neutrinos have
never clustered and their mass gives a scale-independent suppression
of the gravitational potential similar to the effect of increasing $w$.
On intermediate scales, there is a scale-dependent suppression of the
lensing power. The additional effect of keeping $\theta_D$ fixed is to
subtract off a small proportion of the suppression that arises when
$\Omega_\Lambda h^2$ is increased so that neutrino mass then gives
a small positive enhancement of lensing power at low $\ell$ and a suppression
for larger $\ell$ (Fig.~\ref{fig:w_on_deriv}).
The different scale dependence of the effects
of neutrino mass and $w$ allows them to be separated if the
lensing potential can be accurately determined. In this way one
can determine \emph{both} $w$ and $\Omega_\nu h^2$ accurately, whereas
neither (nor any combination) can be determined accurately from the
unlensed CMB alone. Forecasts for such constraints were given
in Ref.~\cite{LEN/KKS03}, which used for the errors on $C_l^{\phi\phi}$
those expected from an application of the quadratic
reconstruction technique of Ref.~\cite{LEN/H+O02}. This exploits the local
scale-scale correlations in the non-Gaussian lensed CMB fields to reconstruct
a (noisy) estimate of the deflection field. Improvements in the reconstruction
of the deflection field may be possible with more optimal
techniques~\cite{LEN/H+S03b}. Reference~\cite{LEN/KKS03} found that
a post-Planck experiment could determine $w$ to within an error of
$0.18$ and detect the mass of a single family of massive neutrinos
if $m_\nu > 0.04\,\text{eV}$.

\begin{figure}
\includegraphics[angle=-90,width = \columnwidth]{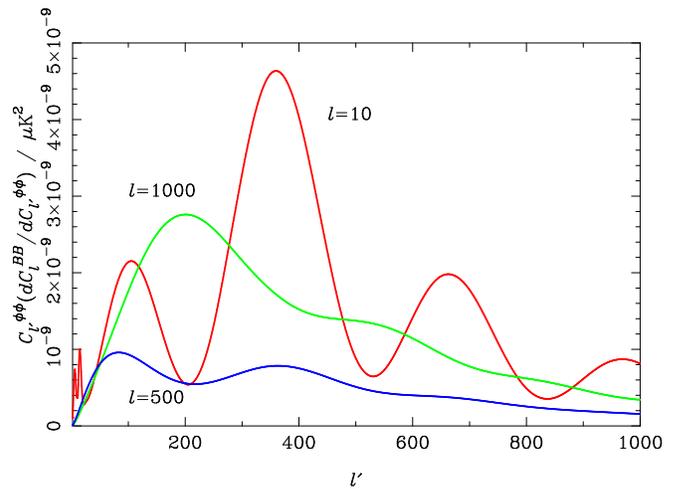}
\caption{(Colour online) Kernels $C_{\ell'}^{\phi\phi} \text{d}C_\ell^{BB} /
\text{d} C_{\ell'}^{\phi\phi}$ for $\ell=10$ (red), $\ell=500$ (blue) and
$\ell=1000$ (green). These give (approximately) the contribution to the
lensed $B$-mode spectrum from lenses at scale $\ell'$.
}
\label{fig:dclbdclphi}
\end{figure}

Figure~\ref{fig:w_on_deriv} shows that $w$ and $\Omega_\nu h^2$ are largely
degenerate with respect to their effect on the $B$-mode power spectrum.
This is because for no $\ell$ is the dominant contribution to
$C_\ell^{BB}$ coming from the largest angle lenses for which there is a
clear difference between the effects of varying $w$ and $\Omega_\nu h^2$;
for $\ell \agt 200$, the effect of varying $w$ or $\Omega_\nu h^2$
is almost degenerate in the lensing potential power spectrum and hence
in $C_\ell^{BB}$.
In Fig.~\ref{fig:dclbdclphi} we plot the derivatives
$C_{\ell'}^{\phi\phi} \text{d}C_\ell^{BB} /
\text{d} C_{\ell'}^{\phi\phi}$ as a function of $\ell'$ for several values
of $\ell$. For any $\ell$, the integral of the kernel over $\ell'$ gives the
lensed $B$-mode power spectrum at that $\ell$ (in the gradient approximation,
Eq.~\ref{eq:born}). Similarly, for the parameter variations considered
here that leave the unlensed $E$-mode power spectrum unchanged, the
convolution of the kernel $C_{\ell'}^{\phi\phi} \text{d}C_\ell^{BB} /
\text{d} C_{\ell'}^{\phi\phi}$
with the fractional change in $C_{\ell'}^{\phi\phi}$ gives the change in
the lensed $B$-mode spectrum. The shape of the kernels in
Fig.~\ref{fig:dclbdclphi} can be understood as follows. For $\ell \alt 200$,
the dominant contribution to the integral in Eq.~(\ref{eq:C^BB})
comes from $\ell' \gg \ell$ since the $E$-modes have very little
power at the large scale $\ell$. In this limit, the lensed $B$-mode spectrum
is approximately white and the kernel reduces to 
\begin{equation}
C_{\ell'}^{\phi\phi} \frac{\text{d}C_\ell^{BB}}{\text{d} C_{\ell'}^{\phi\phi}}
\approx \frac{1}{4\pi} \ell'^5 C_{\ell'}^{\phi\phi} C_{\ell'}^{\tilde{E}
\tilde{E}},
\end{equation}
for $\ell \ll \ell'$. For $\ell$ close to 1000, around the peak of the
$E$-mode spectrum, the dominant contribution to $C_\ell^{BB}$ is from
larger scale lenses and the kernel is of the form
\begin{equation}
C_{\ell'}^{\phi\phi} \frac{\text{d}C_\ell^{BB}}{\text{d} C_{\ell'}^{\phi\phi}}
\approx \frac{1}{4\pi} \ell'^5 C_{\ell'}^{\phi\phi}
\langle C_{\ell}^{\tilde{E} \tilde{E}} \rangle_{\ell'},
\end{equation}
for $\ell > \ell'$. Here,
$\langle C_{\ell}^{\tilde{E} \tilde{E}} \rangle_{\ell'}$ is a smoothed
version of $C_{\ell}^{\tilde{E} \tilde{E}}$ where the smoothing is with
a bimodal kernel of total width $2\ell'$. If $\ell' \ll 200$ the smoothing
has little effect and $\langle C_{\ell}^{\tilde{E} \tilde{E}} \rangle_{\ell'}
\approx C_{\ell}^{\tilde{E} \tilde{E}}$. Finally, for $\ell \agt 5000$
(not shown in Fig.~\ref{fig:w_on_deriv}), the $B$-mode power arises from
lenses at the same scale across which the
CMB may be approximated by a gradient~\cite{LEN/Zal00}. In this limit, the
kernel is roughly
\begin{equation}
C_{\ell'}^{\phi\phi} \frac{\text{d}C_\ell^{BB}}{\text{d} C_{\ell'}^{\phi\phi}}
\sim \frac{1}{4} \ell^2 C_\ell^{\phi\phi} \left(\frac{1}{2\pi}
\int \frac{\text{d}L}{L} {L}^4 C_{L}^{\tilde{E}\tilde{E}}\right) \delta_{\ell
\ell'},
\end{equation}
where the factor in brackets is the mean-squared gradient of the polarization.

Since $w$ and $\Omega_\nu h^2$ are essentially
degenerate in the $B$-mode power spectrum, the consequence of this
is that, if we use only the lensed $B$-mode power spectrum and
do not consider higher-order statistics, there is a tight
degeneracy between the two parameters.
This is still an improvement over
what can be achieved from the unlensed CMB spectra alone which give
no constraints on either parameter, but clearly falls short of what would
be achieved if the lensing potential could be reconstructed directly.
While this is a clear motivation for developing further such reconstruction
techniques, for the reasons mentioned in Sec.~\ref{sec:intro}, it will
still be worthwhile to perform an initial power-spectrum-based analysis.
Furthermore, external data on e.g.\ the Hubble parameter, will assist
in breaking the degeneracy between $w$ and $\Omega_\nu h^2$ in a
power-spectrum-only analysis. For this reason, we now consider in more detail
how to analyse future $B$-mode power spectrum data taking careful account
of the dependencies between power on different scales.

%
%

\section{Analysis assuming Gaussianity}
\label{sec:gaussian}
We begin by illustrating the effect that non-Gaussianity has on estimated
parameter constraints by (wrongly) analysing simulated lensed maps
with the likelihood function appropriate for Gaussian fields. This
extends the work of Ref.~\cite{LEN/SHK04} who showed that the Fisher
estimate of the error on the amplitude of the lensed $B$-mode spectrum
is under-estimated if Gaussianity is assumed.

We used the publicly-available
LensPix\footnote{\protect \url{http://cosmologist.info/lenspix}}
code~\cite{LEN/Lew05}
(which is based on a modified version of
{\sc Healp}ix 1.2\footnote{\protect \url{http://healpix.jpl.nasa.gov}})
to create simulations of the lensed CMB on the full sky. So that the
effects of non-Gaussianity could be clearly seen we considered the
idealised full-sky case with no noise or foreground sources. The lensed map
was calculated directly at the pixel centres of the deflection map
from the spherical multipoles of the 
unlensed polarization.
This avoids introducing any
unwanted bias due to the effects of pixelization error which may arise
when using the alternative interpolation method in the LensPix code, but is
much slower and limited the number of simulations that we could produce.
The sky was generated using the {\sc Healp}ix resolution parameter
$N_\mathrm{side} = 1024$ (giving 3.4-arcmin pixels)
with $\ell_\mathrm{max} = 2048$ to avoid aliasing. The theoretical power
spectra used in the analysis were calculated also 
using $\ell_\mathrm{max} = 2048$. This leads to 
a lensed $B$-mode power spectrum that lacks power particularly
on smaller scales, but our analysis is
self-consistent.

We restricted the analysis to constraining only the values of $w$
and the tensor/scalar ratio $r$, within a flat dark-energy CDM
cosmology with no massive neutrinos. The strong degeneracy between
$w$ and $\Omega_\nu h^2$ means that it was not useful to vary both
parameters, and omitting massive neutrinos significantly speeds up the
computation of the theoretical power spectra. The models used had
scale-invariant curvature fluctuations ($n_s = 1$) with no running.
The spectral index of tensor fluctuations was set
to $n_t = -r/8$, in accordance with predictions of slow-roll inflation
(see, e.g.\ Ref.~\cite{COS/Lid++97} and references therein).
The values of the physical baryon and CDM densities,
$\Omega_b h^2$ and $\Omega_c h^2$, were held fixed along with
$\theta_D$, so that $\Omega_\Lambda$ and $h$ are derived parameters.
Two fiducial models were used, both with $w=-1$, but the second with a much
lower value of $r$ to see if the non-Gaussianity had a significant effect on
the estimated values of $r$.  Model A had a high value of $r = 0.4$
(close to current upper limits~\cite{COS/Sel++05})
with $\Omega_b h^2 = 0.024$, $\Omega_c h^2 = 0.111$, $H_0= 0.66\,
\text{km\,s}^{-1}\,\text{Mpc}^{-1}$, 
the amplitude of scalar fluctuations on $0.05 \, \text{Mpc}^{-1}$ scales
$A_s = 2.6 \times 10^{-9}$ and optical depth $\tau = 0.2$. Model B used
$r = 0.01$, $\Omega_b h^2 = 0.0234$, $\Omega_c h^2 = 0.111$,
$H_0 = 73\,\text{km\,s}^{-1}\,\text{Mpc}^{-1} $, 
$A_s = 2.3 \times 10^{-9}$ and $\tau = 0.14$. (The value of $A_s e^{-2\tau}$
was the same for both models.)

\begin{figure}
\includegraphics[angle = 270, width = \columnwidth]{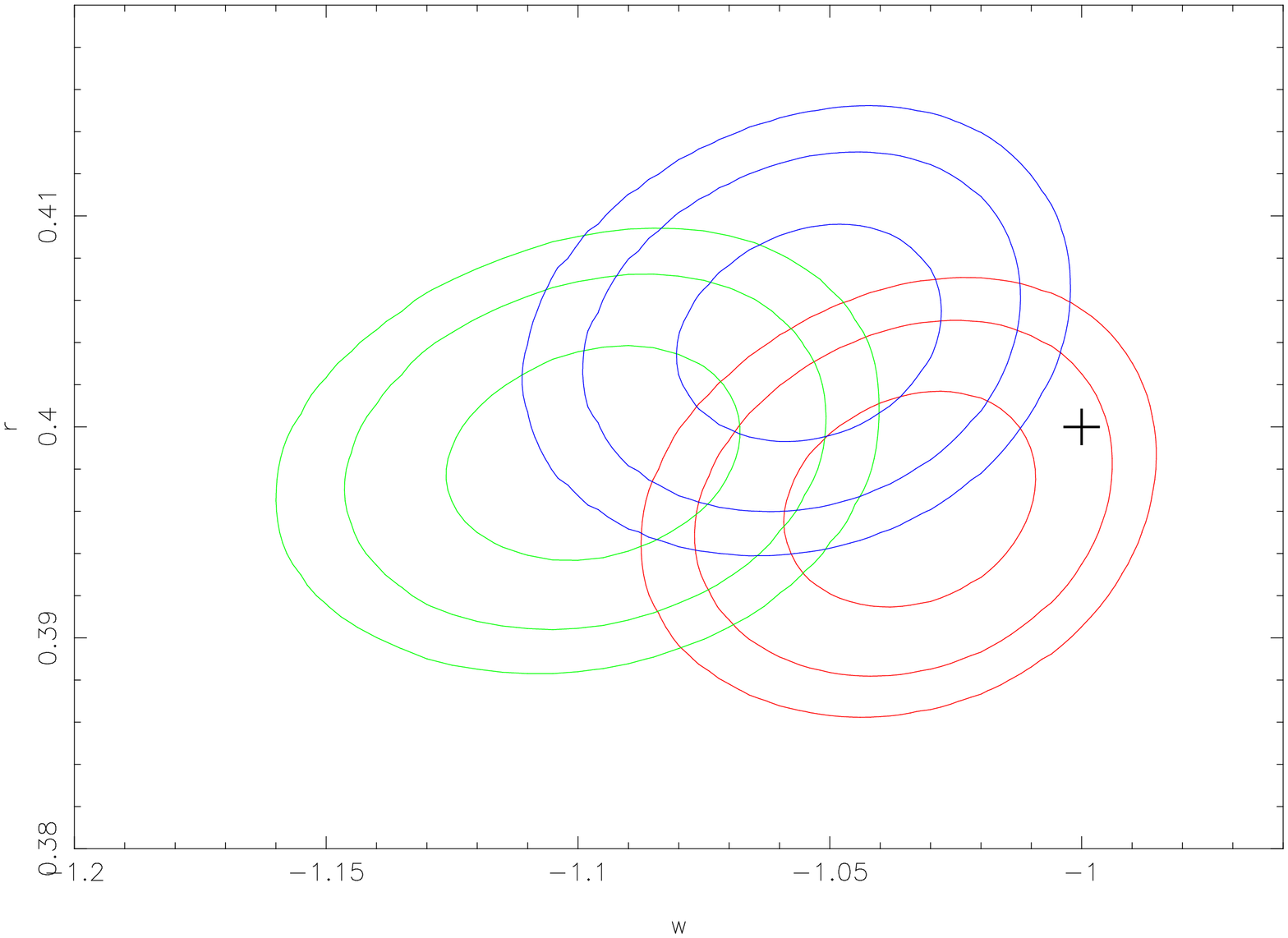}
\includegraphics[angle = 270, width = \columnwidth]{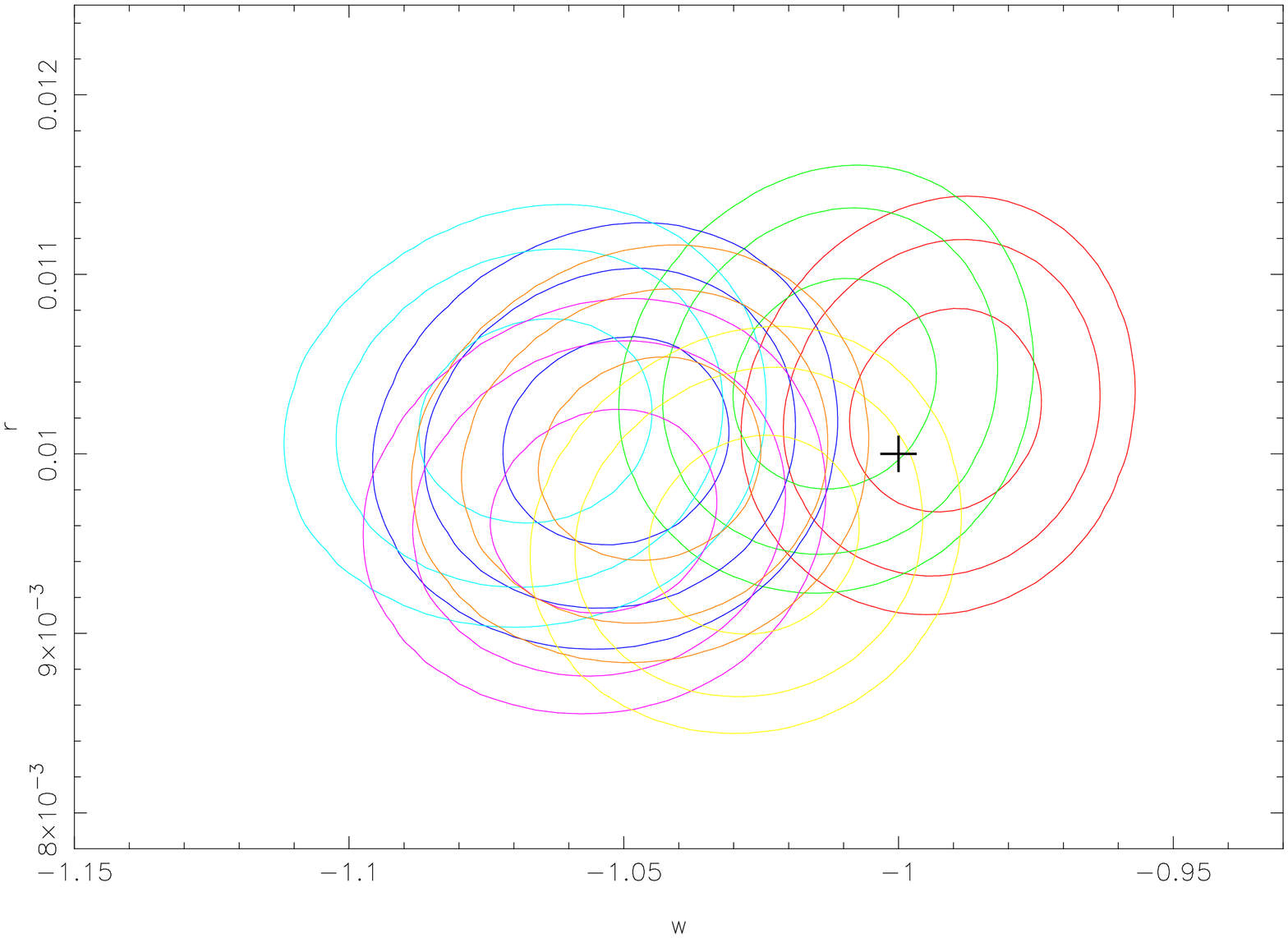}
\caption{Parameter constraints obtained from full-sky lensed simulations
using the $B$-mode power spectrum only and assuming Gaussianity. Contours
are shown at the 67\%, 95\% and 99\% confidence limits. Top: Model-A 
simulations. Bottom: Model-B simulations. The parameters in the fiducial
model are marked with a cross.}
\label{fig:gauss_2d}
\end{figure}

We extracted the measured power spectrum $\hat{C}^{BB}_\ell$ of the simulations
and constrained $r$ and $w$ with the likelihood appropriate
to full-sky, noise-free observations of Gaussian fields:
\begin{equation}
P(\hat{\bm{C}}| \bm{C}) \propto \prod_\ell
\hat{C}_\ell^{\frac{2\ell-1}{2}} C_\ell^{-\frac{2\ell+1}{2}} 
\exp \left\{ 
-\frac{(2 \ell + 1) \hat{C}_\ell}{2 C_\ell} 
\right \},
\label{eq:gauss-like}
\end{equation}
where $C_\ell$ is the theoretical power spectrum.
The normalisation of the likelihood function with respect to
$\hat{C}_\ell$ is usually dropped since in practice it is a fixed quantity
when constraining parameters.
We assumed flat priors on the values of $w$ and $r$.

Figure~\ref{fig:gauss_2d} shows the parameter constraints obtained from three
Model-A simulations and seven Model-B simulations using the likelihood
in Eq.~(\ref{eq:gauss-like}).
It can be seen that the constraints are, in most cases, inconsistent with
the input value of $w = -1$. However, the marginalised constraints on $r$
are in accordance with the input values. There appears to be a bias
towards low values of $w$, and the spread of the maximum-likelihood values
indicates that the constraints on $w$ are too tight. The latter is consistent
with the findings of the Fisher analysis in Ref.~\cite{LEN/SHK04}.

%
%

\section{Non-Gaussian covariance}
\label{sec:cov}

Equation (\ref{eq:C^BB}) shows that the power in lensed $B$-modes on any particular
scale arises from the power on a range of scales in $E$ and $\phi$. As can be
seen in Fig.~\ref{fig:E_B_phi}, the power in the lensing deflection field peaks
at large scales, meaning that a significant fraction of the 
power in $B$-modes on scales around the peak is generated by
large-scale lenses (see also Fig.~\ref{fig:dclbdclphi}).
A particular mode in the lensing deflection field will lens many
different $E$-modes to generate $B$-modes on a range of scales. The power in these
$B$-modes will be related to the power in the single mode of the deflection field and so the measured power in $B$-modes is significantly
correlated between scales \cite{LEN/SHK04}.
In order to account for this non-Gaussianity in our
analysis, we need to calculate the covariance between the measured
$\hat{C}_\ell^{BB}$. The covariance can be expressed as a sum of the
Gaussian and non-Gaussian parts:
\begin{equation}
\text{Cov}(\hat{C}^{BB}_{\ell_i}, \hat{C}^{BB}_{\ell_j}) =
S_{ij} = S_{ij}^{\text{G}} + S_{ij}^{\text{NG}},
\end{equation}
where the Gaussian part arises from the non-connected part of the
$B$-mode four-point function and on the full sky is given by
\begin{equation}
S_{ij}^{\text{G}} = \delta_{ij} \frac{2(C_{\ell_i}^{BB})^2}{2\ell+1}.
\label{eq:gausscov}
\end{equation}
Lensing has only a very small effect on any primordial $B$-modes
from gravitational waves so we can treat these as an independent
Gaussian field; their power spectrum then only enters the Gaussian
part of the covariance matrix. Similar comments would apply to Gaussian
instrument noise.

We can calculate the non-Gaussian part of the covariance matrix
on the flat sky using the gradient approximation Eq.~(\ref{eq:grad-approx}).
The gradient approximation is not suitable for accurate calculation
of the lensed power spectra \citep{LEN/Lew05}. In addition,
the flat-sky approximation introduces per-cent level errors in
the lensed $B$-mode spectrum on all scales~\cite{LEN/C+L05}. Fortunately,
the errors from these approximations tend to cancel above $\ell \sim 400$
and the lensed $B$-mode spectrum can be computed to $\sim 1\%$ accuracy
on such scales using Eq.~(\ref{eq:C^BB}). We
anticipate that the non-Gaussian components of the covariance matrix
computed with the gradient and flat-sky approximations
will have a similar level of accuracy.

We thus calculate the non-Gaussian part of the covariance matrix within
the flat-sky approximation using the bandpower
expression given in Ref.~\cite{LEN/SHK04}:
\begin{eqnarray}
&& S^{\text{NG}}_{ij} = 
\frac{2}{A\alpha_i \alpha_j} 
\int_{\ell \in i} d^2 \bm{\ell}_i
\int_{\ell \in j} d^2 \bm{\ell}_j
\int \frac{d^2 \bm{L}}{(2\pi)^2} \times
\nonumber \\ &&
\left\{
W^2(\bm{\ell}_i, \bm{\ell}_i - \bm{L})
W^2(\bm{\ell}_j, \bm{\ell}_j - \bm{L})
C^{\tilde{E}\tilde{E}}_{|\bm{\ell}_i - \bm{L}|}
C^{\tilde{E}\tilde{E}}_{|\bm{\ell}_j - \bm{L}|}
(C^{\phi\phi}_L)^2 
\right. 
\nonumber \\ &&
+ \;
W^2(\bm{\ell}_i, \bm{L}) W^2(\bm{\ell}_j, \bm{L}) 
(C^{\tilde{E}\tilde{E}}_L)^2 C^{\phi\phi}_{|\bm{\ell}_i - \bm{L}|}
C^{\phi\phi}_{|\bm{\ell}_j - \bm{L}|} 
\nonumber \\ &&
+ \;
W(\bm{\ell}_i, \bm{\ell}_i - \bm{L})
W(-\bm{\ell}_i, \bm{\ell}_j - \bm{L})
W(\bm{\ell}_j, \bm{\ell}_j - \bm{L}) 
\nonumber \\ && \left.
W(-\bm{\ell}_j, \bm{\ell}_i - \bm{L})
C^{\tilde{E}\tilde{E}}_{|\bm{\ell}_i - \bm{L}|}
C^{\tilde{E}\tilde{E}}_{|\bm{\ell}_j - \bm{L}|}
C^{\phi\phi}_L
C^{\phi\phi}_{|\bm{\ell}_i + \bm{\ell}_j - \bm{L}|}
 \right\},
\label{eq:ng_cov}
\end{eqnarray}
where $A$ is the area of sky observed, $\alpha_i = \int_{\ell \in i} d^2 \bm{\ell}_i$
and similarly for $\alpha_j$, and $W$ is given by Eq.~(\ref{eq:W}).
This expression assumes periodic boundary conditions both to remove
the complications due to $E$-$B$ mixing and the additional correlations
between the \emph{observed} Fourier modes due to the geometry. For
surveys of a few-hundred $\text{deg}^2$, the former should be negligible
except on the largest scales, and the latter can be dealt with by choosing
wide enough bands. Neither of these is an issue for the full-sky
observations we consider here, and in this case we can set $A=4\pi$ and
use bands of width $\Delta \ell = 1$.
This means that we can simplify this integral
somewhat, since for a fixed length of $\bm{\ell}_i = \ell_i$
and $\bm{\ell}_j = \ell_j$ the value of the integrand is
dependent only on the angle between $\bm{\ell}_i$ and $\bm{\ell}_j$
and their moduli. Therefore we can
take out the integral over $\bm{\ell}_j$ and instead fix
$\bm{\ell}_j = (\ell_j, 0)$. We can also simplify the 
integral over $\bm{\ell}_i$ to a one-dimensional integral
in $\theta$. We can summarise this as
\begin{equation}
\frac{2}{A\alpha_i \alpha_j} 
\int_{\ell \in i} d^2 \bm{\ell}_i
\int_{\ell \in j} d^2 \bm{\ell}_j
\rightarrow
\frac{1}{4\pi^2} \int_{0}^{2\pi} d \theta_i.
\end{equation}
\begin{figure}
\centering
\includegraphics[width = \columnwidth]{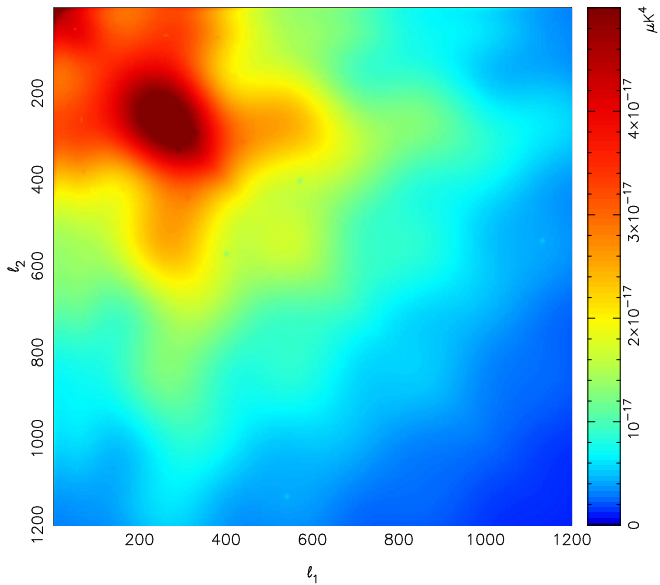}
\caption{(Colour online) Illustration of the non-Gaussian component of covariance matrix
$\mathrm{Cov}(\hat{C}^{BB}_{\ell_1}, \hat{C}^{BB}_{\ell_2})$. The maximum value
at the peak is $5\times 10^{-16}\mu$K$^4$ but the scale has been set
to illustrate the structure in the rest of the matrix more clearly.}
\label{fig:cov_matrix}
\end{figure}

Figure~\ref{fig:cov_matrix}
illustrates the structure of the non-Gaussian component of the
covariance matrix as calculated for Model A. It can be seen that the correlation is present
across all scales and is not confined to the region near the diagonal.
Although the non-Gaussian covariance between any two $\hat{C}_\ell^{BB}$
is small compared to the Gaussian covariance of the diagonal
elements ($\sim 3 \times 10^{-12}\, \mu\text{K}^4 / \ell$ for $\ell
\alt 300$), the fact that the non-Gaussian covariance spans a wide range of
scales means that its net effect is very significant.

It is computationally expensive (taking several hours of CPU time) 
to re-calculate the covariance matrix for each set of model parameters
as would be required in parameter estimation.
However, when considering variations in those parameters that are not
well constrained by the primary anisotropies ($w$ and $\Omega_\nu h^2$ here),
the main effect on the
lensed $B$-mode power spectrum is an overall scaling in amplitude. For this
reason, we can approximate the covariance as 
\begin{equation}
S_{ij} = 
S_{ij}^{\text{G}}
+ \frac{C_{\ell_i}}{C_{\ell_i}^{\mathrm{fid}}}
\frac{C_{\ell_j}}{C_{\ell_j}^{\mathrm{fid}}}
S_{ij}^{\text{NG,\,fid}},
\label{eq:s_ij}
\end{equation}
where $C_\ell$ is the lensing contribution to the $B$-mode power spectrum
and $C_\ell^{\text{fid}}$ is the same in the fiducial model.
This is similar to the scheme suggested in Ref.~\cite{LEN/SHK04}.

\begin{figure}
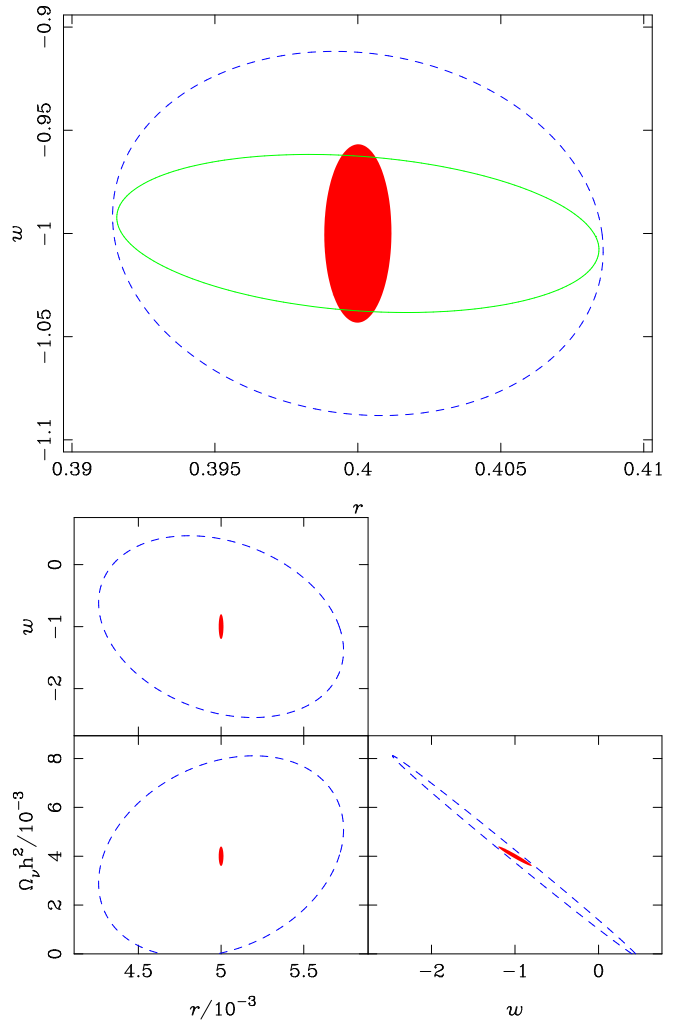

\includegraphics[angle=270,width=\columnwidth]{model1_NGB_ALL_BO}
\includegraphics[angle=270,width=\columnwidth]{model2_NGB_ALL}
\caption{(Colour online) 95\% confidence limits from a Fisher analysis.
Top: Model A. Bottom: constraints for a model with same parameters as Model A
except $r=0.005$ and $\Omega_\nu h^2 = 0.004$ (which we also allow to vary
in the analysis).
Blue dashed contours: limits if the lensed $B$-mode
power spectrum is used, with the correct covariance matrix.
Green solid line (top plot only): limits
using the $B$-mode power spectrum and incorrectly assuming the
data are Gaussian. Red filled area: theoretical limits if all unlensed power
spectra are used including the power spectrum of the lensing deflection field.}
\label{fig:fisher}
\end{figure}

We can calculate the theoretically-optimum constraints on parameters
$\theta_i$ by performing a Fisher analysis. For Gaussian fields, the Fisher
information matrix reduces exactly to~\cite{POL/Z+S97}
\begin{equation}
F_{ij} = \sum_{\ell = 2}^{\ell_\mathrm{max}} \sum_{X,Y}
\frac{\partial C_\ell^{X}}{\partial \theta_i} \,
\mathrm{Cov}^{-1}(C^X_\ell, C^Y_\ell) \,
\frac{\partial C_\ell^{Y}}{\partial \theta_j},
\label{eq:fisher_mult}
\end{equation}
where $X$ and $Y$ can be $TT$, $EE$, $TE$ or $BB$ and 
$\mathrm{Cov}^{-1}(C^X_\ell, C^Y_\ell)$ is the inverse
covariance matrix of the various measured power spectra.
If we wish to consider the information that can be obtained
from the lensed $B$-mode power spectrum, we need to compute the Fisher
matrix $-\langle \partial^2 \ln P/ \partial \theta_i \partial \theta_j
\rangle$ where $P=P(\hat{C}_\ell^{BB}|\bm{\theta})$. If we can approximate
this as Gaussian in the measured power spectrum, then provided the number
of independent degrees of freedom is much larger than
unity\footnote{It is in just this limit of a large number of degrees of
freedom that the likelihood will approach a Gaussian.}
we find
\begin{equation}
F_{ij} \approx \sum_{\ell \ell'}
\frac{\partial C_\ell^{BB}}{\partial \theta_i} \,
S^{-1}_{\ell \ell'} \,
\frac{\partial C_{\ell'}^{BB}}{\partial \theta_j},
\label{eq:fisher_BB}
\end{equation}
where $C_\ell^{BB}$ is the total $B$-mode power spectrum.

The tightest parameter constraints achievable under any circumstances
from CMB data alone are those obtained from the power spectra of the
unlensed, Gaussian, temperature and polarization fields together with
the lensing deflection field. Some approximation to this situation may
be achievable using the full likelihood function for the lensed CMB
fields~\cite{LEN/H+S03,LEN/H+S03b}, or, with larger errors, from quadratic
reconstruction techniques~\cite{LEN/Hu01,LEN/H+O02}.
In the top plot of Fig.~\ref{fig:fisher} we show the constraints obtained
from a Fisher analysis of Model A, if all parameter values are fixed except
$r$ and $w$. We consider three cases: using only the lensed $B$-mode
spectrum with the correct covariance matrix; the same but (wrongly) using only
the Gaussian part of the covariance matrix; and using all of the unlensed
CMB fields and the lensing deflection field (treated as Gaussian). In all
cases we only kept modes up to $\ell=1200$. In the latter case, we include
the unlensed CMB temperature and electric polarization though these have
very little impact since the parameters we allow to vary are chosen
to be nearly degenerate with respect to these fields. 
If the non-Gaussianity is neglected the constraints on $w$ from the
$B$-mode power spectrum alone are slightly tighter than those obtainable
using all of the unlensed fields, which shows immediately that we are
making a false assumption~\cite{LEN/Hu02}.
It can be seen that the errors on $r$ are almost unaffected
by the non-Gaussianity whereas the errors on $w$ are significantly affected.
However the presence of the lensed $B$-modes does significantly
increase the errors on $r$~\cite{POL/K+S02,POL/KSK02},
so if the gravitational wave amplitude is small
($r \alt 0.01$) it will be desirable to attempt reconstruction of the
unlensed CMB in order to detect the signal.

The degeneracy between $w$ and $\Omega_\nu h^2$ that is present when only
the $B$-mode power spectrum is used can be clearly seen in the bottom
plot of Fig.~\ref{fig:fisher}. Here, $r$, $w$ and $\Omega_\nu h^2$
are allowed to vary around a fiducial model with the same parameters
as Model A except for a smaller $r=0.005$ and massive neutrinos
with $\Omega_\nu h^2=0.004$. It should be recognised that a Fisher analysis,
based on the derivatives of the power spectra around the fiducial model
values, is good at highlighting degeneracies but does not give accurate
error contours for degenerate parameters \cite{DA/Eis++99}. The actual
degeneracy between $w$ and $\Omega_\nu h^2$ follows a curve rather than
a straight line. The degeneracy is broken by using
the power spectrum of the lensing deflection field, although even then
the variables are still correlated to a lesser extent. In the unlensed,
foreground-free ideal case, the errors on $r$ are proportional to $r$
and so there is no lower limit to the value of $r$ that can be detected.
In practice foreground residuals (and other systematic effects)
are expected to pose a significant challenge
and may well be the factor that limits the constraints that can
ultimately be achieved.
%

%
%

\section{Non-Gaussian Likelihood}
\label{sec:ng-like}

Having shown clearly that the non-Gaussianity of lensed $B$-modes cannot
be ignored, we now address the question of how it may correctly be taken into
account during parameter estimation from measured power spectra.
To do this, we need to employ a likelihood
function which includes information about the dependencies between
the power on different scales. In principle,
the exact likelihood can be found from that for the lensed
fields~\cite{LEN/H+S03,LEN/H+S03b}, but this would be very complicated,
and such an analysis would probably be no simpler than an optimal
analysis with the fields themselves (which would avoid the lossy compression
to power spectra). Instead we proceed heuristically, modelling the
likelihood as a function of the lensed spectrum and its covariance
only. This necessarily assumes that the connected moments of the
lensed field can be ignored above the four-point level, or, more
correctly, that they can be approximated in a hierarchical manner by the
lower moments. We only consider noise-free analysis on the full sky here, or,
in Sec.~\ref{sec:new-like}, on a periodic flat sky. Since our approach
is based on the measured power spectrum, the additional
(geometric) complications from the survey geometry and those
from instrument noise can be dealt with by standard techniques (see
e.g.~\cite{DA/Cha??}).

A suitable likelihood function must reduce to a good approximation to
the true likelihood
in the limit that the non-Gaussian covariance
tends to zero, Eq.~(\ref{eq:gauss-like}).
Note that it is the dependence of this function
on the cosmological parameters $\bm{\theta}$ that is of interest.
The coupling of power between different scales by non-Gaussianity
has a similar effect to the geometric coupling between scales that arises
in observations of Gaussian fields over only part of the sky. A number
of approximate likelihoods have been suggested for this latter problem,
and a simple strategy for analysing lensed $B$-mode spectra
might be to replace the (geometric) covariance matrix in these
approximations by the non-Gaussian covariance matrix.

A possible first approximation, a likelihood function
that is Gaussian in $\hat{C}_\ell$, is known to produce biased
parameter constraints from Gaussian fields~\cite{DA/BJK00}.
A better approximation is the log-normal
distribution in which the likelihood is Gaussian in the log of the
power. This was originally developed for the case where
the peak of the likelihood (considered as a function of the
theory $C_\ell$\footnote{For Gaussian fields, the parameters
$\bm{\theta}$ only enter the likelihood through the power spectrum.})
and its curvature there are known~\cite{DA/BJK00}, but
it can easily be tailored to our problem by replacing the modal
$C_\ell$ by the measured $\hat{C}_\ell$, and the (inverse) curvature
by the theoretical covariance matrix. (This likelihood was
denoted by $\mathcal{L}_{\text{LN}}'$ in Ref.~\cite{DA/Ver++03}, with
the prime distinguishing it from the original formulation in
Ref.~\cite{DA/BJK00}; we make the same distinction in
Eq.~\ref{eq:WMAP} below.) The log-normal likelihood
has been employed in the analysis of several different CMB data sets,
e.g.~\cite{DA/G++03, CMB/Mac++05, DA/P++02, DA/Bal++00, VSA/R++03}.
However, this likelihood is also biased; the level of bias is
acceptable for many data sets but becomes significant when using
full-sky data with a low noise level. The first-year WMAP data were
analysed using a likelihood function which is a weighted combination
of the Gaussian and log-normal likelihoods \cite{DA/Ver++03}:
\begin{equation}
\ln P(\hat{\bm{C}}|\bm{\theta}) = 
\frac{1}{3}\ln P_\mathrm{Gauss}(\hat{\bm{C}}|\bm{\theta})
 + \frac{2}{3}\ln P'_\mathrm{LN}(\hat{\bm{C}}|\bm{\theta}).
\label{eq:WMAP}
\end{equation}
This is a significantly better approximation to the exact likelihood in the
limit of no non-Gaussianity; see Fig.~\ref{fig:compare_likelihoods} of
Appendix~\ref{ap:newlike} for comparisons.

\begin{figure}
\centering
\includegraphics[angle = 270, width = \columnwidth]{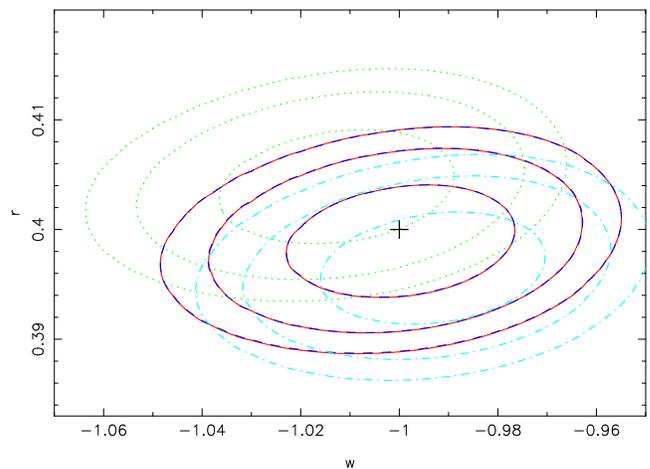}
\caption{(Colour online) Likelihood contours at 67, 95 and 99\% confidence limits
from analysis of a Gaussian simulation of the CMB made using the lensed power spectrum.
Red solid contours: using the exact
likelihood function. Green dotted contours: using the Gaussian approximation to the likelihood
function, showing the significant difference from the exact likelihood.
Light blue dot-dashed contours: using the log-normal
distribution, which gives a smaller, and opposite, shift.
Dark-blue dashed contours: using the WMAP
likelihood function, showing almost indistinguishable results from the exact 
likelihood. The fiducial model values are marked with a cross.}
\label{fig:gauss_sim}
\end{figure}

To illustrate the level of
bias of the different likelihood functions when estimating $r$ and
$w$, we created a Gaussian simulation of the CMB using the lensed
$B$-mode power spectrum, and analysed the resulting measured 
power spectrum. The results are shown in Fig.~\ref{fig:gauss_sim}.
It can be seen that the constraints obtained when the WMAP
likelihood function is employed are virtually indistinguishable
from those obtained using the exact likelihood given by Eq. (\ref{eq:gauss-like}), 
although the Gaussian and log-normal approximations do lead to significant
differences.

\begin{figure}
\centering
\includegraphics[angle = 270, width = \columnwidth]{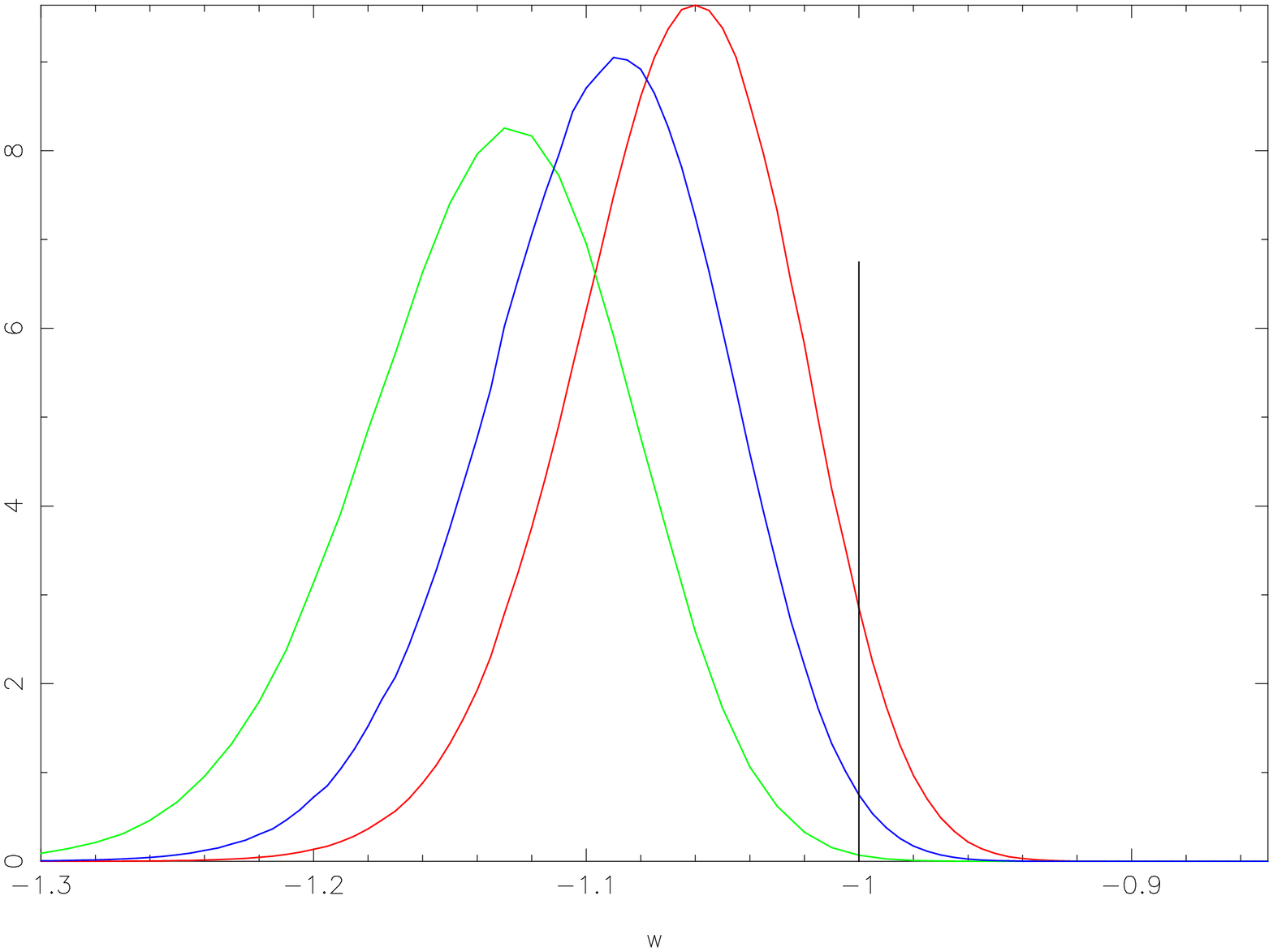}
\includegraphics[angle = 270, width = \columnwidth]{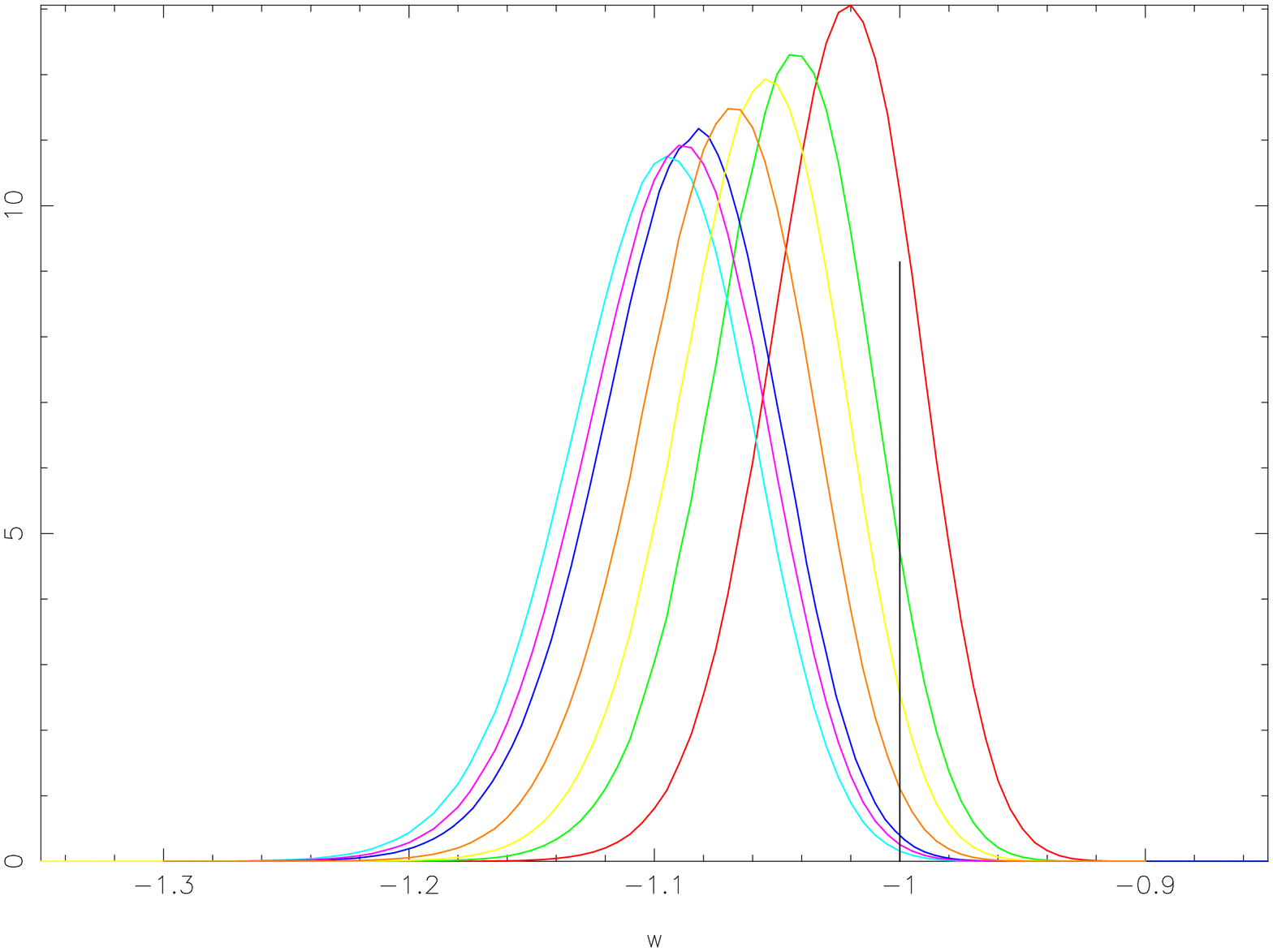}
\caption{Marginalised distributions of $w$ obtained for three Model-A simulations (top) and seven
Model-B simulations (bottom), using the WMAP likelihood function.}
\label{fig:nglike_w}
\end{figure}

Figure~\ref{fig:nglike_w} shows the marginalised distribution of $w$ obtained
when the WMAP likelihood function is used to estimate parameters from
the ten lensing simulations of Fig.~\ref{fig:gauss_2d}. The non-Gaussian
covariance matrix was included with the prescription in
Eq.~(\ref{eq:s_ij}).
Although from the spread
of the maximum-likelihood values it seems that the width of the distributions
are a good reflection of the uncertainties in the parameters, there is a 
significant bias towards low values of $w$, which means that, for most of the
simulations, the parameter estimates are inconsistent with the input values.
The marginalised distributions of $r$ are very similar to those obtained
when Eq. (\ref{eq:gauss-like}) was applied to the same simulations.
This indicates that the non-Gaussian contribution to the covariance is less
significant on large scales, and hence the presence of non-Gaussianity
does not introduce a bias, or affect the errors, on $r$. This is understandable
since the signal which contains the information about $r$ is Gaussian to a 
high level.
%
%
%
\section{New likelihood function}
\label{sec:new-like}

In the previous section we showed that
none of the existing likelihood approximations in the literature
appear suitable for the analysis of future cosmic-variance limited,
lensed $B$-mode data. For this reason we have re-examined the issue
of likelihood approximations in CMB analysis; our findings can be found
in Appendix~\ref{ap:newlike}. The main result is a new likelihood function
that appears to out-perform existing approximations. Here we
summarise this new likelihood function and demonstrate that it produces
accurate parameter constraints on simulated lensed data.

To begin, it is worth first reviewing the underlying issues that arise when
constraining parameters from non-Gaussian fields.
For ideal, full-sky observations, in the absence of non-Gaussianity, 
the probability of the observed fields $\bm{d}$ given
the set of cosmological parameters $\bm{\theta}$ can be written as a
function of the theoretical power spectrum $C_\ell(\bm{\theta})$ and
the measured power spectrum $\hat{C}_\ell$:
\begin{equation}
P(\bm{d}|\bm{\theta}) = P(\bm{d}|C_\ell)=
f(\hat{C}_\ell) P(\hat{C}_\ell | C_\ell),
\end{equation}
where $f(\hat{C}_\ell)$ is a function of the measured power spectrum only.
It follows that $\hat{C}_\ell$ are sufficient statistics for parameter
estimation and hence there is no loss of information in compressing
the data to the measured power spectrum. For non-Gaussian
fields, even for ideal observations, the measured power is no longer a
sufficient statistic and the likelihood may not just be a function
of the theoretical power spectra of the fields.
Nevertheless, we can regard the measured power spectrum as a form of lossy
compression of the data, and the relevant likelihood is then the
sampling distribution $P(\hat{C}_\ell|\bm{\theta})$.

In Appendix~\ref{ap:newlike} we derive a new likelihood
by approximating the properly-normalised distribution
$P(\hat{C}_\ell|\bm{\theta})$ as
Gaussian in some function of the $\hat{C}_\ell$. The likelihood takes the form
\begin{equation}
\ln P(\hat{C}_\ell|\bm{\theta}) \approx \ln A -\frac{1}{2}
\sum_{\ell \ell'} M^{-1}_{\ell \ell'} (\hat{x}_\ell -\mu_\ell)
(\hat{x}_{\ell'} -\mu_{\ell'}),
\label{eq:new-like}
\end{equation}
where 
\begin{eqnarray}
\hat{x}_\ell &=& \hat{C}_\ell^{1/3} \\
\mu_\ell &=& \left(\frac{2\ell-1}{2\ell+1}C_\ell \right)^{1/3},
\end{eqnarray}
and
\begin{equation}
M^{-1}_{\ell \ell'} = 3 C_\ell^{2/3} \left( \frac{2\ell-1}{2\ell+1}\right)^{1/6}
S^{-1}_{\ell \ell'}  \; 
3 C_{\ell'}^{2/3} \left( \frac{2\ell'-1}{2\ell'+1}\right)^{1/6},
\end{equation}
where $S_{\ell \ell'}$ is the covariance matrix of the measured
$\hat{C}_\ell$ at parameters $\bm{\theta}$. The normalisation
is
\begin{equation}
A^{-1} \propto \sqrt{\text{det} M_{\ell \ell'}} \prod_\ell \mu_\ell^2,
\end{equation}
which we approximate here by $A \propto \prod_{\ell} 1/C_\ell$.
Note that the normalisation depends on the cosmological parameters
and including this dependence is important to get an accurate approximation
for the $\bm{\theta}$-dependence of the likelihood in the tails.

\begin{figure*}
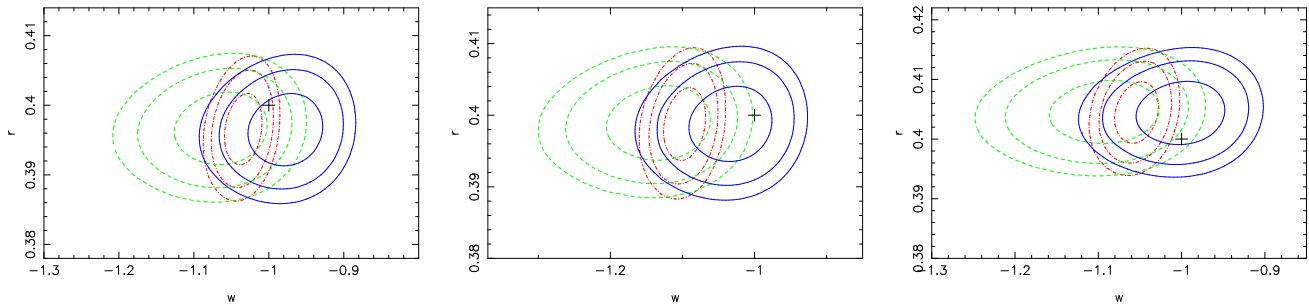

\includegraphics[angle = -90, width = 0.67\columnwidth]{map1_compare_like}
\includegraphics[angle = -90, width = 0.67\columnwidth]{map2_compare_like}
\includegraphics[angle = -90, width = 0.67\columnwidth]{map3_compare_like}
\caption{(Colour online) A comparison of the parameter constraints obtained
for the three Model-A simulations, using different
likelihood functions, with contours shown at 67, 95 and 99\% confidence
limits. Red dot-dash contours: constraints obtained assuming Gaussianity.
Green dashed contours: using WMAP likelihood function. Blue solid
contours: using new likelihood function.
The fiducial model value is shown with a cross.}
\label{fig:compare_like}
\end{figure*}

In Appendix~\ref{ap:newlike} we show that this new likelihood is considerably
more accurate than other approximations in current use when tested
on full-sky, Gaussian data when the number of degrees of freedom is low
(i.e.\ low $\ell$). When applied to Gaussian simulations, we find
parameter constraints very similar to those shown for the WMAP (and exact)
likelihood functions in Fig.~\ref{fig:gauss_sim}. However, the new likelihood
produces very different results when applied to non-Gaussian lensed 
simulations with the non-Gaussian covariance employed, as shown in
Fig.~\ref{fig:compare_like}. It can be seen here that for all three Model-A
simulations the parameters constraints obtained are consistent with the values
used in the simulations, unlike those obtained when the WMAP likelihood
function is used. The results from the Model-B simulations are similar.
In all cases the marginalised constraints
on $r$ are nearly identical to those obtained when Gaussianity is assumed,
which indicates that
the non-Gaussianity does not noticeably affect the estimated value of $r$,
although the presence of the lensed power spectrum itself does influence the
bounds we can place on $r$. The parameter constraints are consistent with
the theoretical errors obtained from the Fisher analysis shown in
Fig.~\ref{fig:fisher}.

In order to test the new likelihood function further, we performed flat-sky
lensed simulations, which can be computed much more rapidly than the full-sky 
simulations. It was found that, with a pixel size of 1.3 arcmin, 
accurate power spectra could be obtained by interpolation provided that a cubic
interpolation method was used (a simple linear interpolation resulted in an inaccurate
lensed power spectrum). Periodic boundary conditions were specified. 
The amplitude of the lensed $B$-mode power spectrum does not vary linearly with
$w$, and for measured power spectra with large uncertainties, this results
in a distribution of $w$ which is highly skewed and stretches far into the
region $w < -1$. To avoid this, we used maps which were $91.7^\circ$
across to reduce the sample variance.
It should be noted that for this size of map approximating the
sky as flat is clearly incorrect; the statistics of the lensed modes
in the simulations will differ at the per-cent level from the
spherical expectations on all
scales~\cite{LEN/C+L05}, and more so for those modes approaching the
survey size.
We performed 150 of these simulations, with Model-A parameters.

The estimated power spectrum of the flat-sky maps is calculated as bandpowers:
\begin{equation}
\hat{\mathscr{C}}_i = \frac{1}{A} 
\frac{\sum_{\bm{\ell}\in i} \ell^2 |B(\bm{\ell})|^2} {2 \pi \sum_{\bm{\ell}\in i}},
\end{equation}
where $B(\bm{\ell})$ is the measured $B$-mode,
the sum is over values of $\bm{\ell}$ that lie in band $i$,
and, recall, $A$ is the area of the sky.
In the limit that $\ell^2 C_\ell / (2\pi)$ is constant within each band,
we find that $\langle \hat{\mathscr{C}}_i \rangle =
\ell^2 C_{\ell_i} / (2\pi)$.
The WMAP likelihood function can straightforwardly be generalised to
work with bandpowers. For the new likelihood function we need to
drop the $(2\ell - 1)/(2\ell +1)$ factors; they are more significant at
low-$\ell$ where the flat-sky approximation does not hold anyway.
The bandpower version of the new likelihood function can then be expressed as
\begin{eqnarray}
P(\hat{\mathscr{C}} | \bm{\theta}) &\propto&
\frac{1}{\prod_B \mathscr{C}_B} \exp \Biggl[ -\frac{1}{2}
\sum_{B B'} \left( \hat{\mathscr{C}}_B^{1/3} - \mathscr{C}_B^{1/3}
\right) 3 \mathscr{C}_B^{2/3}  \nonumber \\
&& \mbox{}
\times \; \mathcal{S}_{B B'}^{-1}
3 \mathscr{C}_{B'}^{2/3}
\left(
\hat{\mathscr{C}}_{B'}^{1/3} - \mathscr{C}_{B'}^{1/3}
\right)
\Biggr],
\end{eqnarray}
where the Gaussian part of the bandpower covariance matrix
$\mathcal{S}_{B_i B_j}$ is given by
\begin{equation}
\mathcal{S}^{\text{G}}_{B_i B_j} = \delta_{ij} \frac{2(2\pi)^2}{A\alpha_i^2} 
\int_{\ell \in i} d^2 \bm{\ell}  
\left(\frac{\ell^2}{2 \pi} C_{\ell}^{BB}\right)^2,
\end{equation}
and the non-Gaussian part can be calculated from the non-Gaussian part of the
the full-sky covariance matrix as:
\begin{equation}
\mathcal{S}^{\text{NG}}_{B_i B_j} =  \frac{4\pi}{A} \frac{1}{\alpha_i \alpha_j} 
\sum_{\ell_i \in i} 2 \pi \ell_i
\sum_{\ell_j \in j} 2 \pi \ell_j
\frac{\ell_i ^2 \ell_j ^2}{(2\pi)^2} S^{\text{NG,\,full}}_{\ell_i \ell_j}.
\end{equation}

Since the posterior distribution of $w$ is skewed, as mentioned above, the
maximum-likelihood value is greater than the mean value. Using the WMAP
likelihood function, both the mean and maximum-likelihood values of $w$ were
significantly biased, with values (averaged over the simulations)
of $-1.075\pm0.009$ and $-1.044\pm0.007$ respectively. The quoted errors
are standard errors in the mean of 150 simulations, and were estimated from
the simulations. For both the mean and maximum-likelihood value, the
bias is comparable to the random error on the measured value; the situation
would worsen if we considered a larger survey area as the random error
would fall but the bias would remain.
Using the new likelihood, the mean of $w$ was $-0.992\pm0.007$ and the
maximum likelihood value $-1.017\pm0.007$, showing that this likelihood
function performs significantly better.
A comparison between the width of the marginalised distributions of
$w$ and the spread of the mean values estimated from each of the flat-sky
simulations gave good agreement, to around one per cent. This shows that
the estimated errors on the value of $w$ are correct once the
non-Gaussian covariance is included.
There was a slight positive bias in the mean
of $r$ of about 0.5 per cent for the new likelihood and around twice
this for the WMAP likelihood. This is about 30\% of the random error
on $r$ for the sample-variance limited observations considered here,
but we expect that the bias is an artifact of the flat-sky
simulations on large scales.

%
%

\section{Conclusions}
\label{sec:conc}
The generation of $B$-mode polarization by weak
gravitational lensing provides a way of measuring cosmological parameters
from the CMB alone that would otherwise be poorly
constrained. We showed that a power spectrum analysis is wasteful,
and in particular suffers from a strong degeneracy between the
equation of state of the dark energy and neutrino masses. A more optimal
analysis that reconstructs the lensing deflection field can break
this degeneracy~\cite{LEN/KKS03}, as can external data. Nevertheless,
compression into the measured power spectrum will be a useful first step
in the analysis of future $B$-mode data as it is likely to be more
robust against real-world effects than more optimal reconstruction
techniques. With this in mind, we have shown that it will be essential
to take the non-Gaussianity of the lens-induced $B$-modes into account
in a high signal-to-noise power-spectrum analysis, not only to avoid
under-estimation of errors~\cite{LEN/SHK04} but also to remove biases in the
parameters that affect lensing; constraints on the gravitational-wave amplitude
are not noticeably affected by non-Gaussianity. Including the
non-Gaussian covariance of the measured $B$-mode power spectrum
in existing likelihood functions that have been used for parameter estimation
from CMB spectra was shown to give biased results on noise-free simulations
of lensed CMB fields. To remedy this we developed a new likelihood
function that is more accurate in the tails of the distribution when
the number of independent modes is low. Due to the non-Gaussian nature
of CMB lensing, the number of such independent modes below a given
multipole is lower than for Gaussian fields. We verified on simulations
that the new likelihood function performs much better than existing
approximations for the particular application considered here.
However, we expect that it will be more generally applicable for
the accurate analysis of the spectra of Gaussian CMB fields on large scales.

\begin{acknowledgments}
The authors would like to thank Antony Lewis for many helpful comments
throughout the course of this work. SS was supported by PPARC and the
Astrophysics Group, Cavendish Laboratory and GR by the Leverhulme
Trust for the duration of this work. AC is supported by the Royal Society.
Some of the results in this paper have been derived using the 
{\sc Healp}ix package~\cite{DA/Gor++05}.
\end{acknowledgments}

\bibliography{sarahreferences}

\appendix

\section{Likelihood approximations}
\label{ap:newlike}

In this appendix we derive the likelihood approximation employed
in Sec.~\ref{sec:new-like}. 
Our strategy for dealing with the non-Gaussianity of lens-induced
$B$-mode data in this paper is heuristic: we include the non-Gaussian
correlation between power on different scales but our choice of
likelihood is motivated by the analysis of Gaussian fields. For this
reason, our new likelihood approximation should also be useful more
generally in CMB analysis. We only consider observations of a single field here
($\hat{C}_\ell^B$); we plan to address
the more general problem of joint analysis of correlated fields
in a future publication.

We develop two approximations motivated by the two ways that CMB power
spectra are usually obtained. In the first, we work with the
Gaussian CMB field $\bm{d}$ directly, and aim to characterise
the probability $P(\bm{d}|C_\ell)$ as a function of the theoretical
power spectrum $C_\ell$ that encodes all of the cosmological information
$\bm{\theta}$. Exploring $P(\bm{d}|C_\ell)$ is computationally expensive
so often the modal value of $C_\ell$ and the curvature (or a Fisher
approximation to it) are found, e.g.\ Ref.~\cite{DA/Bon++98}, and
used in an analytic approximation to the $C_\ell$-dependence
of $P(\bm{d}|C_\ell)$. Complications due to, for example, the survey
geometry, are accounted for only through their impact on the curvature
matrix and modal $C_\ell$. We construct an approximate likelihood
by looking for variables $x_\ell(C_\ell)$ in which the exact likelihood
is accurately represented by a Gaussian, i.e.\ we approximate the
$C_\ell$-dependence of $\ln P(\bm{d}|C_\ell)$ as
\begin{equation}
\ln P(\bm{d}|C_\ell) \approx -\frac{1}{2} \sum_{\ell \ell'} M^{-1}_{\ell \ell'}
(x_\ell -\mu_\ell) (x_{\ell'} -\mu_{\ell'}),
\label{eq:app1}
\end{equation}
up to an irrelevant constant. If this approximation is to peak at the
correct place, we require $\mu_\ell = x_\ell (C_{\ell,{\text{ml}}})$,
where $C_{\ell,{\text{ml}}}$ is the modal (or maximum-likelihood) value
of $C_\ell$. Similarly, if the curvature at the peak is $-\clf_{\ell\ell'}$,
we must have
\begin{equation}
M^{-1}_{\ell \ell'} = \frac{\text{d} C_\ell}{\text{d} x_\ell}
\frac{\text{d} C_{\ell'}}{\text{d} x_{\ell'}} \clf_{\ell\ell'},
\label{eq:app2}
\end{equation}
where the derivatives are evaluated at the peak. To constrain the variable
change $x_\ell(C_\ell)$, we examine the third derivatives of
$\ln P(\bm{d}|C_\ell)$ with respect to the $x_\ell$:
\begin{eqnarray}
\frac{\partial^3 \ln P}{\partial x_\ell \partial x_{\ell'}\partial x_{\ell''}}
&=& \delta_{\ell \ell'}\delta_{\ell \ell''} \frac{\partial \ln P}{\partial
C_\ell} \frac{\text{d}^3C_\ell}{\text{d}x_\ell^3} \nonumber \\
&&\mbox{}
\hspace{-0.2\columnwidth}
+ \left(\delta_{\ell \ell'}  \frac{\partial^2 \ln P}{\partial
C_\ell \partial C_{\ell''}}
\frac{\text{d}^2 C_\ell}{\text{d}x_\ell^2}
\frac{\text{d}C_{\ell''}}{\text{d}x_{\ell''}} + \text{cyclic perms.}
\right) \nonumber \\
&&\mbox{} + \frac{\partial^3 \ln P}{\partial C_\ell \partial C_{\ell'}
\partial C_{\ell''}}
\frac{\text{d}C_{\ell}}{\text{d}x_{\ell}}
\frac{\text{d}C_{\ell'}}{\text{d}x_{\ell'}}
\frac{\text{d}C_{\ell''}}{\text{d}x_{\ell''}},
\label{eq:app3}
\end{eqnarray}
where have imposed that $x_\ell$ is only a function of $C_\ell$ at the same
$\ell$. Setting the right-hand side exactly equal to zero is, in general,
inconsistent. Instead, we obtain a tractable and consistent problem
if we demand that it vanish everywhere when we replace
the derivatives of $P$ on the right with their expectation values
evaluated with the likelihood appropriate to full-sky, noise-free
observations~\cite{DA/BJK00}:
\begin{equation}
\ln P(\bm{d}|C_\ell) = -\sum_\ell (\ell+1/2)(\hat{C}_\ell / C_\ell +
\ln C_\ell),
\label{eq:app4}
\end{equation}
where $\hat{C}_\ell$ is the measured $C_\ell$.
Evaluating the expectation
values of the derivatives of this likelihood (over data from an ensemble with
power $C_\ell$), we find from Eq.~(\ref{eq:app3}) that we must have
\begin{equation}
4 (\text{d} C_\ell / \text{d} x_\ell)^2 = 3 C_\ell \text{d}^2 C_\ell
/ \text{d} x_\ell^2 .
\label{eq:app5}
\end{equation}
Note that the full-sky number of degrees of freedom $2\ell+1$ present
in the derivatives cancels in this equation so we expect this
procedure to remain valid for observations covering only a fraction
$f_{\text{sky}}$ of the sky.
Equation~(\ref{eq:app5}) is solved by $x_\ell = C_\ell^{-1/3}$, and gives an
approximation for the $C_\ell$-dependence of $\ln P(\bm{d}|C_\ell)$ of the form
\begin{eqnarray}
\ln P(\bm{d}|C_\ell) &=& - \frac{1}{2} \sum_{\ell \ell'} [
3 C_{\ell,\text{ml}}^{4/3} \clf_{\ell \ell'} 3 C_{\ell',\text{ml}}^{4/3}
\nonumber \\
&&\mbox{} \hspace{-0.1\columnwidth}
\times (C_\ell^{-1/3} - C_{\ell,\text{ml}}^{-1/3})
(C_{\ell'}^{-1/3} - C_{\ell',\text{ml}}^{-1/3})],
\label{eq:app6}
\end{eqnarray}
up to an irrelevant constant. The essential difference with
the log-normal approximation developed in Ref.~\cite{DA/BJK00} is that
they demand that the expectation of the curvature of the likelihood be
constant (when approximated by Eq.~\ref{eq:app4}) whereas we force
the expectation value of the derivative of the curvature to be zero. These are
not equivalent since the latter preserves the distinction between
$C_{\ell,\text{ml}}$ and $C_\ell$ until after the derivative of the curvature is
taken while the former does not.

To include instrument noise, we repeat the above steps but now include
isotropic noise in the full-sky likelihood in Eq.~(\ref{eq:app4}). The
effect in our approximate likelihood, Eq.~(\ref{eq:app6}), is to
add a noise offset $N_\ell$ to $C_{\ell,\text{ml}}$ and $C_\ell$ in
the right-hand side. Following Ref.~\cite{DA/BJK00}, we suggest
determining $N_\ell$ by fitting the model
\begin{equation}
\mathcal{F}^{-1}_{\ell \ell} \approx \frac{2}{(2\ell+1)f_{\text{sky}}}
(C_{\ell,\text{ml}}+N_\ell)^2,
\end{equation}
to the diagonal elements of the inverse curvature at the peak.

The second approximation we consider is that which is used in
Sec.~\ref{sec:new-like}. In this case, we first compress the data
down to a measured power spectrum $\hat{C}_\ell$ and then ask what is the
sampling distribution for the $\hat{C}_\ell$ given the cosmological model
$\bm{\theta}$? We then use the $\bm{\theta}$-dependence of this
probability as the likelihood in parameter estimation~\cite{DA/Wan++01}.
For noise-free observations of Gaussian fields over the full sky,
the sampling distribution is given by Eq.~(\ref{eq:gauss-like}) which
on taking logs becomes
\begin{eqnarray}
\ln P(\hat{C}_\ell|C_\ell) &=& - \sum_{\ell} \Big( \frac{2\ell+1}{2}
\frac{\hat{C}_\ell}{C_\ell} + \frac{2\ell+1}{2} \ln C_\ell \nonumber \\
&&\mbox{} \phantom{\sum_\ell} - \frac{2\ell-1}{2} \ln \hat{C}_\ell \Big),
\label{eq:app7}
\end{eqnarray}
up to a constant that is independent of $\hat{C}_\ell$ and the
theoretical spectrum $C_\ell$.
This approach is more natural for the non-Gaussian problem considered in the
body of this paper since there we perform (lossy) compression down to
$\hat{C}_\ell^{BB}$, and the power spectrum of the lensed $B$-modes does not
fully characterise their statistics. We now look for a variable
transformation $\hat{x}_\ell(\hat{C}_\ell)$ in which the sampling distribution
$P(\hat{C}_\ell|\bm{\theta})$ is approximately Gaussian:
\begin{equation}
\ln P(\hat{C}_\ell|\bm{\theta}) \approx \ln A -\frac{1}{2}
\sum_{\ell \ell'} M^{-1}_{\ell \ell'} (\hat{x}_\ell -\mu_\ell)
(\hat{x}_{\ell'} -\mu_{\ell'}),
\label{eq:app8}
\end{equation}
where $A$ is a $\hat{C}_\ell$-independent (but cosmology-dependent)
normalisation.
We assume that we can calculate the mean $\langle \hat{C}_\ell \rangle$
and covariance $S_{\ell \ell'}$ of the $\hat{C}_\ell$ given the
cosmological model. Here we shall assume that $\hat{C}_\ell$ is constructed
to be unbiased so $\langle \hat{C}_\ell \rangle = C_\ell$ although this
assumption can easily be dropped. Since the mean and covariance of
$P(\hat{C}_\ell|\bm{\theta})$ are specified rather than the modal
value of $\hat{C}_\ell$ and the curvature there, we postpone their
determination until we have determined the functional form of
$\hat{x}_\ell$. This is obtained by examining the third
derivatives of the sampling distribution with respect to $\hat{C}_\ell$.
These are given by Eq.~(\ref{eq:app3}) but with $\hat{C}_\ell$ and
$\hat{x}_\ell$ replacing $C_\ell$ and $x_\ell$. Again we approximate
the derivatives with respect to $\hat{C}_\ell$,
but this time with those derived from
the distribution in Eq.~(\ref{eq:app7}). The second and third
derivatives are independent of $C_\ell$ and we make no further approximation
with these. We ignore the first derivative term since it is small
near the peak, so, on setting the (approximate) third derivative of
$\ln P$ with respect to $\hat{x}_\ell$ to zero, we find
\begin{equation}
2 (\text{d} \hat{C}_\ell / \text{d} \hat{x}_\ell)^2 = 3
\hat{C}_\ell \text{d}^2 \hat{C}_\ell / \text{d} \hat{x}_\ell^2 .
\label{eq:app9}
\end{equation}
This is solved by $\hat{x}_\ell = \hat{C}_\ell^{1/3}$. We could
now determine $\mu_\ell$ and $M_{\ell\ell'}$ by demanding that the
approximate distribution gives the correct mean and covariance for
$\hat{C}_\ell$. A simpler method, which we adopt here, is to
relate $\mu_\ell$ and $M_{\ell\ell'}$ to the (known) mean and covariance
using Eq.~(\ref{eq:app7}) as a guide. For that ideal distribution,
the peak is at $[(2\ell-1)/(2\ell+1)]\langle \hat{C}_\ell \rangle$
and the curvature at the peak is related to (minus) the inverse variance by
a factor $[(2\ell+1)/(2\ell-1)]$. This motivates setting
\begin{eqnarray}
\mu_\ell &=& \left(\frac{2\ell-1}{2\ell+1}C_\ell\right)^{1/3}
\label{eq:app10} \\
M^{-1}_{\ell\ell'} &=& \sqrt{\frac{2\ell+1}{2\ell-1}} \frac{\text{d}
\hat{C}_\ell}{\text{d}\hat{x}_\ell} S^{-1}_{\ell\ell'}
\sqrt{\frac{2\ell'+1}{2\ell'-1}} \frac{\text{d}
\hat{C}_{\ell'}}{\text{d}\hat{x}_{\ell'}}, \label{eq:app11}
\end{eqnarray}
where the derivatives are taken at the peak where $\hat{x}_\ell = \mu_\ell$.
If we use this to approximate the exact result Eq.~(\ref{eq:app7}) and
evaluate the mean and standard deviation of $\hat{C}_\ell$, we find
\begin{eqnarray}
\frac{\langle \hat{C}_\ell \rangle}{C_\ell} &=& 1+ \frac{2}{27} \delta^2
+ \dots \label{eq:app12} \\
\sqrt{\frac{2\ell+1}{2C_\ell^2}} \Delta \hat{C}_\ell &=&
1+ \frac{1}{18} \delta + \dots , \label{eq:app13}
\end{eqnarray}
where $\delta \equiv (\ell-1/2)^{-1}$. At $\ell=2$ the fractional
error on the mean and standard deviation are 3\% and 4\% respectively;
by $\ell=10$ these have dropped to $0.1\%$ and $0.6\%$.
It remains to fix the normalisation. Exponentiating Eq.~(\ref{eq:app8})
and integrating with the approximation $\text{d} \hat{C}_\ell \approx
3 \mu_\ell^2 \text{d} \hat{x}_\ell$, we find that
\begin{equation}
A^{-1} \propto \sqrt{\text{det} M_{\ell \ell'}} \prod_\ell \mu_\ell^2,
\label{eq:app14}
\end{equation}
where the proportionality constant is independent of the cosmological
parameters $\bm{\theta}$. In the text we further approximate the normalisation
by $A \propto 1/ \prod_\ell C_\ell$.

To generalise our approximate sampling distribution to include instrument
noise, we assume that the measured power spectrum has had a noise
bias $N_\ell$ removed, so that $\langle \hat{C}_\ell \rangle = C_\ell$
still holds. The modifications we suggest are then
$\hat{x}_\ell = (\hat{C}_\ell + N_\ell)^{1/3}$ and
replacing $C_\ell$ by $C_\ell + N_\ell$ in $\mu_\ell$. 

\begin{figure*}
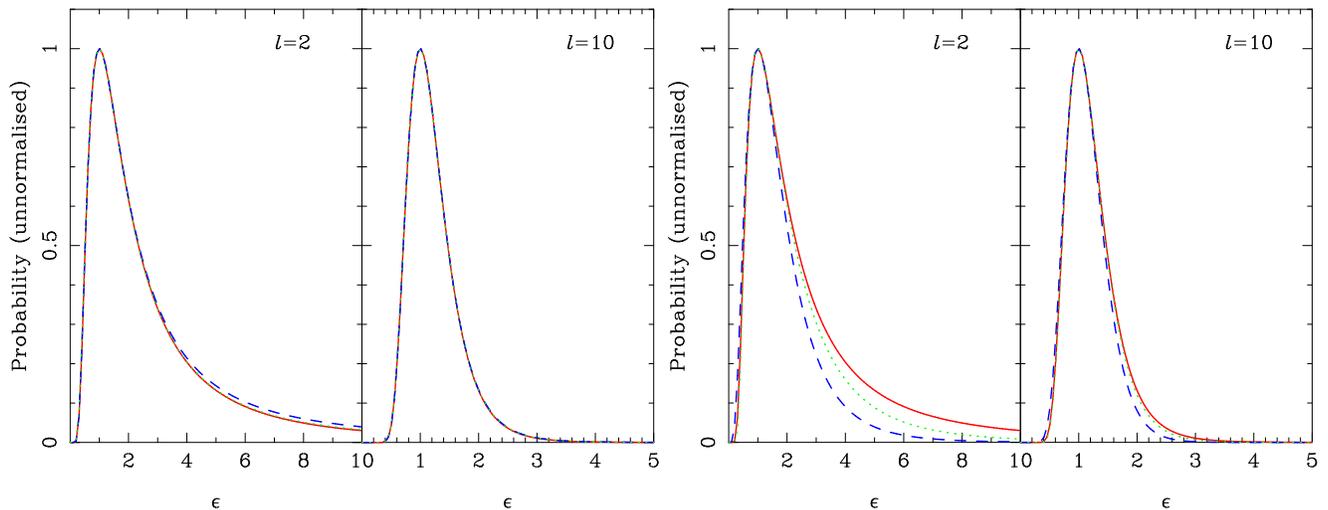

\includegraphics[angle=-90,width = \columnwidth]{exact_ml_pcl_likelihood.ps}
\includegraphics[angle=-90,width = \columnwidth]{exact_wmap_logn_likelihood.ps}
\caption{(Colour online) Comparison of likelihood approximations for
$C_\ell$ for noise-free, full-sky observations of Gaussian fields. Left:
the exact distribution (red, solid), Eq.~(\ref{eq:app4}); our new result
(blue, dashed) in Eq.~(\ref{eq:app6}) that assumes the peak position
and curvature there; and our alternative result (green, dotted) in
Eq.~(\ref{eq:app8}) that works from the sampling distribution of
the measured power spectrum, for $\ell=2$ (left panel) and $\ell=10$ (right
panel).
Note that the green dotted line is barely distinguishable from the exact
distribution.
Right: the log-normal distribution (blue, dashed); the WMAP likelihood
(green, dotted) used in the analysis of the first-year data; and
the exact likelihood (red, solid). In all cases $\epsilon \equiv C_\ell/
\hat{C}_\ell$, where $\hat{C}_\ell$ is the measured $C_\ell$; $\hat{C}_\ell$
is also the maximum likelihood $C_\ell$ for the ideal conditions assumed.
}
\label{fig:compare_likelihoods}
\end{figure*}

In Fig.~\ref{fig:compare_likelihoods} we compare our two new likelihood
approximations with the exact distribution, and also with the log-normal
distribution proposed in Ref.~\cite{DA/BJK00} and the likelihood
used in the first-year WMAP analysis~\cite{DA/Ver++03}, for the case of
noise-free, full-sky observations of Gaussian fields. In this case, the
measured power spectrum $\hat{C}_\ell$ equals the maximum likelihood
spectrum $C_{\ell,\text{ml}}$, and working from the exact sampling distribution
for $\hat{C}_\ell$ produces the same distribution for $C_\ell$ as working
directly from the exact $P(\bm{d}|C_\ell)$. Our new approximations are
seen to be more accurate than the existing approximations, and particularly
so when the number of degrees of freedom ($2\ell + 1$ in this example)
is small. Our second approximation, based on the sampling distribution
of the measured $C_\ell$, is seen to be more accurate in the tail of
the distribution than our first approximation that starts from the peak
position and curvature. This appears to be because of the
non-perturbative way in which the theoretical $C_\ell$ appear in the
normalisation of the sampling distribution.

\end{document}